\documentclass[letterpaper,english,aps,reprint]{revtex4-1}
\usepackage[T1]{fontenc}
\usepackage[latin9]{inputenc}
\usepackage[active]{srcltx}
\usepackage{mathrsfs}
\usepackage{amsmath}
\usepackage{amssymb}
\usepackage{graphicx}

\makeatletter

\usepackage{epstopdf}

\makeatother

\usepackage{babel}
\begin{document}
\preprint{APS/123-QED}
\title{Quantum Mean-Force Kinetic Theory: General Formulation and Application
to Electron-Ion Transport in Warm Dense Matter}
\author{Shane Rightley}
\email{shane-rightley@uiowa.edu}

\affiliation{Department of Physics and Astronomy~\\
 University of Iowa}
\author{Scott D. Baalrud}
\email{scott-baalrud@uiowa.edu}

\affiliation{Department of Physics and Astronomy~\\
 University of Iowa}
\date{\today}
\begin{abstract}
We present an approach to extend plasma transport theory into the
Warm Dense Matter (WDM) regime characterized by moderate Coulomb coupling
and electron degeneracy. It is based on a recently proposed closure
of the BBGKY hierarchy that expands in terms of the departure of correlations
from their equilibrium value, rather than in terms of the strength
of correlations. This kinetic equation contains modifications to the
collision term in addition to a second term that models the non-ideal
contributions to the equation of state. An explicit collision operator
is derived in the semiclassical limit that is similar to that of the
Uehling-Uhlenbeck equation, but where scattering is mediated by the
potential of mean force (PMF). As a demonstration, we use this collision
integral to evaluate temperature and momentum relaxation rates in
dense plasmas. We obtain degeneracy- and coupling-dependent ``Coulomb
integrals'' that take the place of $\ln\Lambda$ in the scattering
rates. We additionally find a novel difference in the way in which
degeneracy influences momentum relaxation in comparison to temperature
relaxation. Finally, we evaluate electron-ion relaxation rates for
the case of warm dense deuterium over a range of density and temperature
spanning the classical to quantum and weak to strong coupling transitions.
Results are compared with the Landau-Spitzer rate and rates obtained
from the quantum Landau-Fokker-Planck equation and Lee-More model.
We find that the models diverge significantly in the degenerate and
moderately coupled regime and attribute this difference to how the
various models treat the physics of Pauli blocking, correlations,
large-angle scattering, and diffraction.
\end{abstract}
\maketitle

\section{\label{sec:introduction}Introduction}

Physical models are often based on expansions in small parameters
about some more simple state. For instance in plasma physics transport
properties are traditionally calculated from an expansion in which
the ratio of the average potential energy to kinetic energy (the Coulomb
coupling parameter $\Gamma=\left\langle U\right\rangle /\left\langle K\right\rangle $)
is taken to be small. In Fermi liquid theory, the coupling may be
strong, but the ratio of the temperature to the Fermi energy (the
inverse degeneracy parameter $\Theta=T/E_{F}$) is taken to be
small. In solid state physics, a rigid or semi-rigid ion lattice provides
a zero-order structure about which to base a perturbation expansion.
Warm Dense Matter (WDM) is defined by the condition that neither the
deviation from an ideal plasma, nor from a zero-temperature Fermi
liquid, nor from a rigid lattice structure, can be taken to be small.
This occurs in experiments involving extreme compression of materials
\citep{Glenzer2016,Riley2018,Mancic2010}, in astrophysics \citep{Redmer2008,Koenig2005},
and along the compression path in inertial confinement fusion (ICF)
experiments \citep{Hu2015}. As a result of the demanding conditions
for theoretical modeling, the description of WDM has been highly reliant
on computational techniques. However, ab initio computation proves
too expensive for many problems in WDM, whereas faster methods often
involve uncontrolled approximations or have uncertain applicability
in WDM conditions. In order to support computational efforts, explore
larger regions of parameter space, and expediently provide data tables
for hydrodynamic simulations, reliable and fast tools for the computation
of transport coefficients in dense plasmas and WDM remain desirable.

In this work, we apply a new expansion parameter to quantum kinetic
theory to develop a model for transport in plasmas subject to electron
degeneracy and Coulomb coupling - thus providing an expansion parameter
suitable for WDM. The model is derived via an extension of a novel
closure \citep{Baalrud2019} of the BBGKY hierarchy to the quantum
domain. Rather than expanding about small density or weak coupling,
the new closure expands in terms of the departure of correlations
from their equilibrium values. In the classical case this model shows
that two particles in the plasma interact via the potential of mean
force, and it has proven successful at predicting ion transport in
both classical and dense plasmas with moderate coupling \citep{Baalrud2013,Baalrud2014,Daligault2016}.
Here, we develop the extension of this theory into the quantum domain
to derive quantum mean force kinetic theory (QMFKT). The modified
BBGKY hierarchy can be closed at any level, which leads to a formal
integrodifferential equation for the reduced Wigner distributions.
The ``free-streaming'' part of the equation in our case contains
a new term associated with the excess pressure; i.e., the model incorporates
non-ideal effects in the equation of state. Meanwhile, the new collision
term contains the effects of correlations, approximated by their equilibrium
limit. Further reduction of the BBGKY hierarchy to a Boltzmann-type
equation is complicated by coupled momentum and position dependence
of the reduced Wigner functions due to the uncertainty principle.
To obtain an immediately useful transport equation without neglecting
diffraction or Pauli blocking, we apply a semi-classical approximation.
The smallness of the electron-ion mass ratio justifies the application
of this approach to electron-ion collisions in plasmas in which the
electron coupling and degeneracy are small to moderate. By closing
the resulting hierarchy at second order, and applying the molecular
chaos approximation, we obtain a convergent quantum collision operator
that is in the form of the Boltzmann-Uehling-Uhlenbeck (BUU) equation,
except where the particles scatter via the potential of mean force
rather than the Coulomb potential. The model in this limit is particularly
suitable for the description of electron-ion relaxation processes
in WDM conditions, and as a demonstration we apply it to the calculation
of relaxation rates in warm dense deuterium.

\begin{figure}
\includegraphics[width=8.6cm]{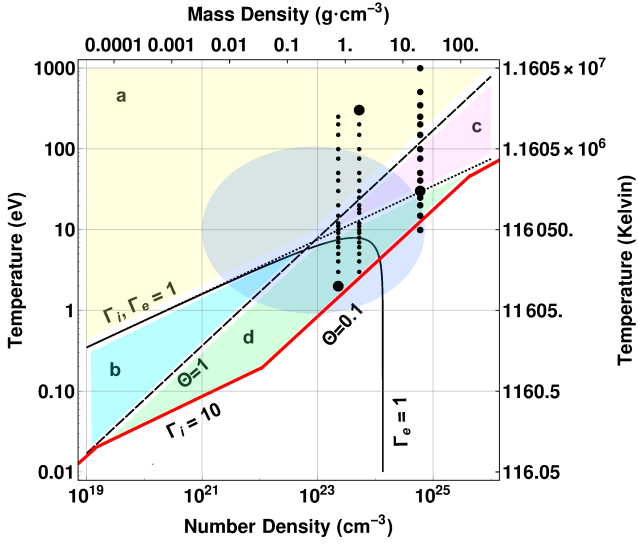}\caption{\label{fig:parregimes}Parameter regimes of plasmas. The solid black
line is the boundary between weak and strong electron coupling $\Gamma_{e}=1$
and turns over due to the electron degeneracy; the dotted lined is
the separation between weak and strong ion coupling $\Gamma_{i}=1$;
and the dashed line is the separation between classical and degenerate
electrons $\Theta=1$. The darker blue oval denotes the sector of
Warm Dense Matter. Region a (yellow) is classical weakly coupled plasma;
region b (light blue) is characterized by classical strong coupling;
region c (pink) by quantum weak coupling, and region d (green) by
both quantum electrons and strongly-coupled ions. We expect QMFKT
to apply to each region a-d. The red line demarcates the region of
validity of plasma-type transport theories; beyond this is the regime
of condensed matter. Dots are points at which the theory is evaluated,
with the bold dots being presented in figure \ref{fig:wide-1}.}
\end{figure}

The approximate regimes in which these different physical processes
are important can be roughly understood in terms of the degeneracy
parameter $\Theta$ and Coulomb coupling parameter $\Gamma$. Because
of degeneracy, the electron Coulomb coupling can differ from that
of the ions even at the same temperature with single ionization. This
is due to the average speed of electrons shifting from the thermal
speed to the Fermi speed as degeneracy increases, a phenomenon that
causes electrons to become increasingly weakly coupled at high density.
This can be seen through the definition $\Gamma=\left\langle U\right\rangle /\left\langle K\right\rangle $
with the statistical averages taken using a Maxwell-Boltzmann distribution
for ions and a Fermi-Dirac distribution for electrons so that
\begin{equation}
\Gamma_{i}=\frac{Z^{2}e^{2}/a}{k_{B}T},
\end{equation}
and
\begin{equation}
\Gamma_{e}=\Gamma_{i}\frac{{\rm Li}_{3/2}\left[-\exp(\mu/k_{{\rm B}}T)\right]}{{\rm Li}_{5/2}\left[-\exp(\mu/k_{{\rm B}}T)\right]},
\end{equation}
where $a=(3/4\pi n)^{1/3}$ is the Wigner-Seitz radius, ${\rm {\rm Li}}$
is the polylogarithm function (closely related to the Fermi integral)
and $\mu$ is the electron chemical potential related to $\Theta$
through the normalization of the Fermi-Dirac distribution \citep{Melrose2010}:
\begin{equation}
-{\rm Li}_{3/2}\left[-\exp(\mu/k_{{\rm B}}T)\right]=\frac{4}{3\sqrt{\pi}}\Theta^{-3/2}.\label{eq:xi_Theta_relation}
\end{equation}
The conditions $\Gamma=1$ and $\Theta=1$ divide the density-temperature
parameter space into multiple regions, as seen in figure \ref{fig:parregimes}.
The regimes can be broken down into (a) classical weak coupling, (b)
classical strong coupling, (c) quantum weak coupling, and (d) classical
strongly coupled ions with degenerate weak or strongly coupled electrons.
WDM exists at the intersection of all of these regions marked by the
blue region, where no small expansion parameter has been available.
Transport in region (a) is well-understood in terms of the Landau-Spitzer
theory, and region (c) has been successfully modeled through quantum
weak-coupling theories such as the quantum Landau-Fokker-Planck equation
\citep{Daligault2016a}. Progress has recently been made extending
classical plasma transport theory into region (b) for $\Gamma\lesssim20$
through use of mean force kinetic theory (MFKT) \citep{Baalrud2013,Baalrud2014,Baalrud2019},
which has also been successfully applied in region (d) for WDM in
the case of ion transport \citep{Daligault2016}. Here, we broaden
this approach to include electron degeneracy in the dynamics.

Existing theoretical methods for predicting transport in WDM typically
fall into the categories of binary collision theories \citep{Daligault2016a,Daligault2018,Gericke2002,Lee1984,Starrett2018},
linear response theories \citep{Daligault2009,Lampe1968,Scullard2018},
and non-equilibrium Green's functions and field-theoretic methods
\citep{Brown2005,Brown2007,Singleton2007,Balzer2013,Bonitz2016}.
In this work we overcome some limits of existing methods by performing
a controlled expansion, and in the quasi-classical limit the resulting
model is physically intuitive and can be evaluated relatively quickly.
The model in this limit is expected to be valid for plasmas with weak
and moderate coupling ($\Gamma\lesssim20$) and classical or moderately
degenerate electrons, when $m_{e}T_{i}\ll m_{i}T_{e}$, and when the
system correlations are well-approximated by their equilibrium limit.
The extension of this region beyond the currently well-described regions
is shown as the green area in figure \ref{fig:parregimes}. For the
case of warm dense deuterium, we evaluate the model and some common
and simple alternatives and discuss the relative importance of the
effects of correlation, large-angle scattering, Pauli blocking, and
diffraction. We use this scheme to select three densities at which
to evaluate the model for warm dense deuterium in order to access
regimes with a different physical process dominating in each.

We begin by deriving our kinetic theory from the quantum Liouville
equation in section \ref{sec:theory}. We introduce the new closure
and, in the limit that electron coupling and degeneracy are small
to moderate, obtain a quasi-classical potential of mean force that
moderates two-body interactions in the plasma. From there we obtain
a quantum kinetic equation, in which the collision operator is similar
to the BUU collision operator, but where correlation effects arise
in the scattering cross section via the potential of mean force. In
section \ref{sec:Relaxation} we apply this to electron-ion momentum
and temperature relaxation, where we obtain degeneracy- and correlation-dependent
``Coulomb integrals'' that replace the traditional Coulomb logarithm.
We compare the relaxation rates predicted from our theory with other
common models in an experimentally relevant parameter regime in section
\ref{sec:discussion}. We conclude and summarize in section \ref{sec:conclusions}.

\section{\label{sec:theory}Quantum Mean Force Kinetic Theory}

Quantum kinetic theory has been developed partly in parallel with
the historical development of plasma kinetic theory and partly as
a result of the demands of advances in solid state physics and high
energy density physics. An overview of the subject can be found in
\citep{Bonitz2016}. Beginning with the quantum Liouville equation
for the Wigner quasi-probability distribution, we will now develop
a closure scheme for the quantum BBGKY hierarchy in which the new
closure parameter quantifies the departure of the correlations in
the system from their equilibrium values.

\subsection{Wigner Functions and the Quantum BBGKY Hierarchy}

Our theory is developed in terms of the many-body Wigner function
$f^{(N)}$, which contains equivalent information to the $N-{\rm body}$
wavefunction, i.e. it completely specifies the state of the system.
The Wigner function evolves according to the quantum Liouville equation
\citep{Imam-Rahajoe1967,Hoffman1965}

\begin{gather*}
\frac{\partial f^{(N)}\left(1,2...N;t\right)}{\partial t}+\sum_{i=1}^{N}\frac{\boldsymbol{p}_{i}}{m_{i}}\cdot\nabla_{i}f^{(N)}\left(1,2...N;t\right)\\
+\text{\ensuremath{\mathscr{O}}}\cdot f^{(N)}\left(1,2...N;t\right)=0
\end{gather*}
with a local potential $U(\boldsymbol{r}_{N})$ which is a function
of the $3N$ coordinates $\boldsymbol{r}_{N}$ of the $N$ particles.
Here the notation $\left(1,2...N\right)$ means the set of all $N$
positions and momenta $\left(\{\boldsymbol{\boldsymbol{r}}_{N}\},\left\{ \boldsymbol{p}_{N}\right\} \right)$
and
\begin{equation}
\text{\ensuremath{\mathscr{O}}}\cdot f^{(N)}=-\frac{2}{\hbar}\sin\left[\frac{\hbar}{2}\frac{\partial^{U}}{\partial\boldsymbol{r}_{N}}\cdot\frac{\partial^{f}}{\partial\boldsymbol{p}_{N}}\right]U(\boldsymbol{r}_{N})f^{(N)}(1,2...N;t),
\end{equation}
in which the superscripts $U$ and $f$ refer to the function on which
the derivatives operate. The derivatives represent $3N-{\rm dimensional}$
gradients, and the $\sin$ operator is defined via its power series.
For a pairwise (e.g. Coulomb) potential $\phi(\boldsymbol{r}_{ij})$
with no external fields,
\begin{equation}
U(\boldsymbol{r}_{N})=\frac{1}{2}\sum_{i=1}^{N}\sum_{j=1,\thinspace j\neq i}^{N}\phi_{ij}(\boldsymbol{r}_{i},\boldsymbol{r}_{j}),
\end{equation}
and
\begin{equation}
\text{\ensuremath{\mathscr{O}}}=-\frac{1}{\hbar}\sum_{i=1}^{N}\sum_{j=1,\thinspace j\neq i}^{N}\sin\left[\frac{\hbar}{2}\frac{\partial^{\phi}}{\partial\boldsymbol{r}_{ij}}\cdot\frac{\partial}{\partial\boldsymbol{p}_{i}}\right]\phi(\boldsymbol{r}_{ij}),
\end{equation}
where $\boldsymbol{r}_{ij}=\boldsymbol{r}_{i}-\boldsymbol{r}_{j}$.
To facilitate comparison with the Liouville equation, we expand the
$\sin$ operator,
\begin{widetext}
\begin{equation}
\left\{ \frac{\partial}{\partial t}+\sum_{i=1}^{N}\left[\frac{\boldsymbol{p}_{i}}{m_{i}}\cdot\frac{\partial}{\partial\boldsymbol{r}_{i}}-\frac{1}{\hbar}\sum_{j=1,\thinspace j\neq i}^{N}\sum_{k=0}^{\infty}\frac{(-1)^{k}}{(2k+1)!}\left(\frac{\hbar}{2}\right)^{2k+1}\frac{\partial^{2k+1}\phi(\boldsymbol{r}_{ij})}{\partial\boldsymbol{r}_{ij}^{k+1}}\cdot\frac{\partial}{\partial\boldsymbol{p}_{i}^{2k+1}}\right]\right\} f^{(N)}\left(1,2...N;t\right)=0\;.
\end{equation}
The classical Liouville equation is obtained by keeping only the first
term of the sum over $k$. To maintain similarity with the classical
case, we write the quantum Liouville equation as
\begin{equation}
\left\{ \frac{\partial}{\partial t}+\sum_{i=1}^{N}\left[L_{i}+\sum_{j=1,\thinspace j\neq i}^{N}L_{ij}^{QC}\right]\right\} f^{(N)}\left(1,2...N;t\right)=0,\label{eq:qbbgky}
\end{equation}
where
\begin{gather}
L_{i}\equiv\frac{\boldsymbol{p}_{i}}{m_{i}}\cdot\frac{\partial}{\partial\boldsymbol{r}_{i}},\label{eq:qLi}
\end{gather}
and
\begin{gather}
L_{ij}^{QC}\equiv-\frac{1}{\hbar}\sum_{k=0}^{\infty}\frac{(-1)^{k}}{(2k+1)!}\left(\frac{\hbar}{2}\right)^{2k+1}\frac{\partial^{2k+1}\phi(\boldsymbol{r}_{ij})}{\partial\boldsymbol{r}_{ij}^{2k+1}}\cdot\frac{\partial}{\partial\boldsymbol{p}_{i}^{2k+1}},\label{eq:qLij}
\end{gather}
is the ``quantum Coulomb'' operator. Equations (\ref{eq:qbbgky})-(\ref{eq:qLij})
are an exact description of the $N-{\rm body}$ problem.
\end{widetext}

As is true classically, full knowledge of the N-body problem is not
necessary to specify the important parameters of the system. Instead,
the $n{\rm 'th}$ reduced Wigner function, obtained by integrating
the N-body Wigner function over some fraction $N-n$ of the particle
phase space
\begin{equation}
f^{(n)}(1,2...,N;t)\equiv\frac{N!}{(N-n)!}\int d\boldsymbol{r}^{(N-n)}d\boldsymbol{p}^{(N-n)}f^{(N)},
\end{equation}
where $d\boldsymbol{x}^{n}=d\boldsymbol{x}_{1}d\boldsymbol{x}_{2}\ldots d\boldsymbol{x}_{n}$,
is sufficient to calculate observables that depend on $n$ or fewer
particle coordinates. For example, the fluid variables number density,
drift velocity, pressure, and energy-density can be determined from
the reduced one-particle Wigner functions.

If the quantum Liouville equation is integrated over the $N-n$ subspace
coordinates, an equation of motion for the $n-{\rm body}$ reduced
Wigner function $f^{(n)}\left(1,2...n;t\right)$ is obtained. Averaging
over the positions and momenta of particles $n$, $n+1$... $N$,
utilizing the symmetry of the Wigner function under exchange of particle
indices, provides the quantum BBGKY hierarchy
\begin{gather}
\left[\frac{\partial}{\partial t}+\sum_{i=1}^{n}\left(L_{i}+\sum_{j=1,\thinspace j\neq i}^{N}L_{ij}^{QC}\right)\right]f^{(n)}(1,2,...n;t)\nonumber \\
=-\sum_{i=1}^{n}\int\int d^{3}r_{n+1}d^{3}p_{n+1}L_{i,n+1}^{QC}f^{(n+1)}(1,2,...n+1;t).\label{eq:bbgky}
\end{gather}
An approximation is needed in order to close the hierarchy and make
calculations using only the first $n$ reduced functions. Typically,
the approximation that is made involves the assumption that $f^{(n+1)}=0$,
implying \textit{weak correlations, }in order to neglect dependence
on the $n+1$ and higher-order distributions. In the following, we
will follow a recent approach that has been applied to the classical
limit \citep{Baalrud2019} and instead make the approximation that
the system is close to equilibrium, without neglecting correlations.

\subsection{Expansion Parameter and Closure Approximation}

Typically, equation (\ref{eq:bbgky}) is closed at a given level $n$
by making some assumption about the unimportance of correlations at
the level $n+1$, thus simplifying or neglecting the $f^{(n+1)}$
dependence on the right-hand side (RHS). However, this approximation
neglects correlations and is not justified in even a weakly-coupled
plasma because screening is important. To obtain a convergent kinetic
equation, we instead re-write the equation in terms of a new expansion
parameter that does not explicitly depend on the strength of the interaction
or the density. We define the relative deviation of the $n+1{\rm -body}$
Wigner function from equilibrium as
\begin{equation}
\frac{\Delta f^{(n+1)}}{f_{0}^{(n+1)}}\equiv\frac{f^{(n+1)}}{f_{0}^{(n+1)}}-\frac{f^{(n)}}{f_{0}^{(n)}}.
\end{equation}
Adding and subtracting the second term on the right hand side of this
definition to the BBGKY equation and rearranging, we express $\Delta f$
as the expansion parameter:
\begin{gather}
\Bigg\{\frac{\partial}{\partial t}+\sum_{i=1}^{n}\Bigg[L_{i}+\sum_{j=1,\thinspace j\neq i}^{n}L_{ij}^{QC}\nonumber \\
+\int\int d^{3}r_{n+1}d^{3}p_{n+1}L_{i,n+1}^{QC}\frac{f_{0}^{(n+1)}}{f_{0}^{(n)}}\Bigg]\Bigg\} f^{(n)}\\
=-\sum_{i=1}^{n}\int\int d^{3}r_{n+1}d^{3}p_{n+1}L_{i,n+1}^{QC}\Delta f^{(n+1)}.\label{eq:qBBGKYNEW}
\end{gather}
The operator $L_{i,n+1}^{QC}$ in the last term on the LHS is understood
to contain derivatives that operate on both the equilibrium Wigner
functions and the $f^{(n)}$ term. Like the original BBGKY equation,
this is an exact description of the plasma. However, the RHS is now
proportional to a quantity which is by definition small for a system
near equilibrium. Closure can then be obtained naturally at any level
$n$ of the hierarchy by taking $\Delta f^{(n+1)}=0$. The remainder
on the LHS and the first $n-1$ equations then describe the effective
dynamics of $n$ particles in the presence of the static influence
of particles $n+1\ldots N$ within the equilibrium reduced Wigner
function $f_{0}^{(n+1)}$ on the LHS.

The relevant operator is the new term on the left hand side
\begin{equation}
\sum_{i=1}^{n}\int\int d^{3}r_{n+1}d^{3}p_{n+1}L_{i,n+1}^{QC}\left(\frac{f_{0}^{(n+1)}}{f_{0}^{(n)}}\right).\label{eq:bbgkyoperator}
\end{equation}
In the classical case this term can be shown to combine with the free-particle
operator $L_{i}$ to become a new two-body interaction operator containing
the potential of mean force \citep{Baalrud2019}. However, the equilibrium
Wigner functions differ from the classical distributions in that they
must encapsulate the uncertainty and Pauli-exclusion principles. Specifically,
the $f_{0}^{(n)}$ do not decouple into independent functions of the
positions and momenta of the particles due to the uncertainty principle.
In the classical limit, the ratio $f_{0}^{(n+1)}/f_{0}^{(n)}$ depends
on the momentum $\boldsymbol{p}_{n+1}$ only through a factor $\exp\boldsymbol{p}_{n+1}$,
and after the integration over $\boldsymbol{p}_{n+1}$ the momentum
derivative in the $L_{i,n+1}$ operator therefore commutes through
this ratio. Without this simplification, it is not clear how to write
a new operator in the form $\nabla W_{ij}\cdot\partial f^{(n)}/\partial p_{ij}$,
and therefore unclear how to explicitly obtain a potential of mean
force. Furthermore, the $1-{\rm particle}$ momentum distribution
for a fermion does not have simple Maxwellian dependence but instead
is a Fermi-Dirac distribution. This complicates the derivation, and
the former even obscures the meaning of the potential of mean force
for a general quantum system due to the dependence on momentum.

Nevertheless, the closed hierarchy represents a formal approximate
solution. Kinetic theories provide closed equations for the single-particle
distribution. This may be obtained through closure at the second hierarchy
equation. Taking $\Delta f^{(3)}=0$, the system is described by the
first level equation
\begin{gather}
\left[\frac{\partial}{\partial t}+L_{1}+\int\int d^{3}r_{2}d^{3}p_{2}L_{12}^{QC}\frac{f_{0}^{(2)}\left(1,2\right)}{f_{0}^{(1)}\left(1\right)}\right]f^{(1)}\label{eq:firstbbgky}\\
=-\int\int d^{3}r_{2}d^{3}p_{2}L_{12}^{QC}\Delta f^{(2)}
\end{gather}
and the second level equation
\begin{gather}
\Bigg\{\frac{\partial}{\partial t}+L_{1}+L_{2}+L_{12}^{QC}+L_{21}^{QC}\nonumber \\
+\int\int d^{3}r_{3}d^{3}p_{3}\left[\left(L_{13}^{QC}+L_{23}^{QC}\right)\frac{f_{0}^{(3)}}{f_{0}^{(2)}}\right]\Bigg\} f^{(2)}=0.\label{eq:secondbbgky}
\end{gather}
Consider the first level equation; typically, the LHS would describe
the free streaming motion of a particle while the RHS encapsulates
its interaction with the rest of the plasma, i.e. the collision operator.
In this new scheme however, the last term on the LHS depends on the
equilibrium correlations of the rest of the plasma. This term is associated
with the non-ideality of the plasma. Taking the momentum moment $\int d\boldsymbol{p}_{1}\boldsymbol{p}_{1}\left\{ \cdots\right\} $,
the gradient of the ideal pressure results from the free-streaming
term $\int d\boldsymbol{p}_{1}\boldsymbol{p}_{1}L_{1}f^{(1)}=\nabla P_{k}$
where $nk_{B}T$ is the ideal component of the pressure while the
new term is
\begin{equation}
\int d\boldsymbol{p}_{1}\boldsymbol{p}_{1}\int\int d\boldsymbol{r}_{2}d\boldsymbol{p}_{2}L_{12}^{QC}\frac{f_{0}^{(2)}\left(1,2\right)}{f_{0}^{(1)}\left(1\right)}.
\end{equation}
In the limit of a Maxwellian distribution and $\hbar\rightarrow0$,
it has been shown \citep{Baalrud2019} that this becomes exactly the
gradient of the excess pressure $p_{\phi}=-(\rho^{2}/6)\int_{0}^{\infty}dr\phi'(r)rg(r)$
in classical mechanics \citep{Hansen2006}, where $r=|\boldsymbol{r}_{1}-\boldsymbol{r}_{2}|$.
Whether such an analytic reduction exists in the quantum case remains
to be seen, but it seems clear that the interpretation can be carried
over.

Considering the second equation of the hierarchy (\ref{eq:secondbbgky}),
this can be seen as an integrodifferential equation for the reduced
Wigner function $f^{(2)}$. Knowledge of $f^{(2)}$ and thus $\Delta f^{(2)}$
would provide the collision operator that would appear on the RHS
of equation (\ref{eq:firstbbgky}). A direct solution of equation
(\ref{eq:secondbbgky}) would allow direct calculation of transport
coefficients for plasmas of arbitrary constituency, strong degeneracy
and with weak to moderate coupling. However, such a general solution
seems unlikely except under conditions in which the operator (\ref{eq:bbgkyoperator})
simplifies. We will for now restrict ourselves to a semiclassical
limit that fulfills these conditions.

\section{Semi-classical Limit}

In order to make further progress, and as a demonstration of the theory,
in addition to the closure approximation $\Delta f^{(3)}\ll f_{0}^{(3)}$
we from this point onward assume a limit in which the operator (\ref{eq:bbgkyoperator})
is well-approximated by its classical limit. We will proceed to justify
this specifically for the case of electron-ion interactions in the
limit that the ions are classical and the electron degeneracy and
coupling are both moderate, but it is also applicable to ion-ion collisions
and semiclassical electron-electron collisions. This can be seen through
application of the semi-classical Wigner-Kirkwood expansion in $\hbar$
for the functions $f_{0}$. The expressions are lengthy and the details
have been reserved for Appendix \ref{sec:wignerkirkwoodappendix}.
Applying the results of Shalitin \citep{Shalitin1973}, we can write
e.g.
\begin{gather}
\frac{f_{0}^{(n+1)}}{f_{0}^{(n)}}\approx\left(\frac{m}{2\pi k_{B}T}\right)\exp\left(-\frac{1}{2mk_{B}T}p_{n+1}^{2}\right)\\
\times\frac{\rho_{n+1}(\boldsymbol{r}_{1},...,\boldsymbol{r}_{n+1})}{\rho_{n}(\boldsymbol{r}_{1},...,\boldsymbol{r}_{n})}\left[1+\left(\frac{Q_{n+1}}{C_{n+1}}-\frac{Q_{n}}{C_{n}}\right)\right]
\end{gather}
where 
\begin{equation}
\rho_{n}\equiv\frac{N!}{(N-1)!}\frac{\int d\boldsymbol{r}^{(N-n)}\exp\left[-V_{N}/k_{B}T\right]}{\int d\boldsymbol{r}^{(N)}\exp\left(-V_{N}/k_{B}T\right)}
\end{equation}
is the $n-{\rm body}$ correlation function with $V_{N}\equiv\sum_{i=1}^{N}\sum_{j>i}^{N}\phi_{ij}$
being the total electrostatic potential energy and where the $Q$
and $C$ functions are integrals of combinations of gradients of the
potential and contain the sole dependence on the momenta of particles
$1$ through $n$; see Appendix \ref{sec:wignerkirkwoodappendix}.
Furthermore, these terms are of second and higher order in the potential,
and are second order in terms of quantum effects. They are thus suppressed
by the classical nature of the ions and lack of strong coupling of
the electrons, a condition that is well-matched in the WDM regime.
We therefore approximate $f_{0}^{(n)}=\rho_{n}(\boldsymbol{r}^{(n)})\Pi_{i}^{n}f_{M}^{(1)}(\boldsymbol{p}_{i})$
so that
\begin{equation}
\frac{f_{0}^{(n+1)}}{f_{0}^{(n)}}\approx\frac{\rho_{n+1}(\boldsymbol{r}_{1},...,\boldsymbol{r}_{n+1})}{\rho_{n}(\boldsymbol{r}_{1},...,\boldsymbol{r}_{n})}f_{M}^{(1)}\left(\boldsymbol{p}_{n+1}\right),\label{eq:wignerapprox}
\end{equation}
where $f_{M}$ is the Maxwellian distribution. This separation of
the position and momentum dependence results in a momentum-independent
potential of mean force, but necessitates restriction of the current
theory to the semiclassical limit. Without this assumption it is unclear
how to interpret the last term on the LHS of equation (\ref{eq:qBBGKYNEW}).
With this approximation, we obtain
\begin{gather}
\frac{1}{\hbar}\sum_{k=0}^{\infty}\left(\frac{\hbar}{2}\right)^{2k+1}\frac{(-1)^{k}}{(2k+1)!}\Bigg\{\sum_{j=1,\thinspace j\neq i}^{n}\frac{\partial^{k}\phi}{\partial\boldsymbol{r}_{ij}^{k}}-\\
\int d\boldsymbol{p}_{n+1}f_{M}^{(1)}(\boldsymbol{p}_{n+1})\int d\boldsymbol{r}_{n+1}\frac{\rho_{n+1}}{\rho_{n}}\frac{\partial^{2k+1}\phi}{\partial\boldsymbol{r}_{i,n+1}^{2k+1}}\Bigg\}\\
=-\sum_{j=1,\thinspace j\neq i}^{n}\frac{1}{\hbar}\sum_{k=0}^{\infty}\frac{(-1)^{k}}{(2k+1)!}\left(\frac{\hbar}{2}\right)^{2k+1}\frac{\partial^{2k+1}W^{(n)}}{\partial\boldsymbol{r}_{i}^{2k+1}}
\end{gather}
where
\begin{gather}
\frac{\partial^{2k+1}W^{(n)}}{\partial\boldsymbol{r}_{i}^{2k+1}}\equiv\sum_{j=1,\thinspace j\neq i}^{n}\frac{\partial^{2k+1}\phi}{\partial\boldsymbol{r}_{ij}^{2k+1}}\nonumber \\
+\int d\boldsymbol{p}_{n+1}f_{M}^{(1)}(\boldsymbol{p}_{n+1})\int d\boldsymbol{r}_{n+1}\frac{\rho_{n+1}}{\rho_{n}}\frac{\partial^{2k+1}\phi}{\partial\boldsymbol{r}_{i,n+1}^{2k+1}}\label{pomf}
\end{gather}
defines what can properly be interpreted as the potential of mean
force $W^{(n)}$. The BBGKY hierarchy can be re-written with the mean-force
potential absorbing the $f_{0}$ terms on the LHS and the Coulomb-force
term
\begin{gather*}
\left\{ \frac{\partial}{\partial_{t}}+\sum_{i=1}^{n}\left[L_{i}+\bar{L}_{i}^{Q}\right]\right\} f^{(n)}=\\
-\sum_{i=1}^{n}\int\int d\boldsymbol{r}_{n+1}d\boldsymbol{p}_{n+1}L_{i,n+1}^{QC}\Delta f^{(n+1)}\left(1,2...n+1\right)
\end{gather*}
where
\begin{equation}
\bar{L}_{i}^{Q}\equiv-\frac{1}{\hbar}\sum_{k=0}^{\infty}\frac{(-1)^{k}}{(2k+1)!}\left(\frac{\hbar}{2}\right)^{2k+1}\frac{\partial^{2k+1}W^{(n)}}{\partial\boldsymbol{r}_{i}^{2k+1}}\cdot\frac{\partial}{\partial\boldsymbol{p}_{i}^{2k+1}}
\end{equation}
now describes $n-{\rm body}$ interactions through the potential of
mean force $W^{(n)}$. This takes the same form as the initial hierarchy
equation, but the force operator on the LHS has been modified and
now depends on the equilibrium Wigner functions at the $n+1$ level.
Finally, we note that this semiclassical limit of the potential of
mean force can be written in terms of the correlation function 
\begin{equation}
g^{(n)}(\boldsymbol{r}_{1},...,\boldsymbol{r_{n}})\equiv\rho^{(n)}(\boldsymbol{r}_{1},...,\boldsymbol{r_{n}})/\Pi_{i=1}^{n}\rho^{(1)}(\boldsymbol{r}_{i})\label{eq:paircorrelation}
\end{equation}
 as
\begin{equation}
W^{(n)}=-k_{B}T\ln g^{(n)},\label{eq:PMFcalc}
\end{equation}
which provides direct connection between the potential of mean force
and the structural properties of the plasma.

\subsection{Derivation of a Boltzmann-Uehling-Uhlenbeck-like Equation}

The theory at this point is written in the form of the original BBGKY
equation, but with the Coulomb interaction operator $L_{ij}^{QC}$
being replaced by the mean-force operator $\bar{L}_{i}^{Q}$. Classically,
the Boltzmann equation is derived by closing at the second level of
the hierarchy. With the new operator $\bar{L}_{i}^{Q}$, equations
(\ref{eq:firstbbgky}-\ref{eq:secondbbgky}) become
\begin{gather}
\left[\frac{\partial}{\partial t}+L_{1}+\int\int d\boldsymbol{r}_{2}d\boldsymbol{p}_{2}L_{12}^{QC}\frac{\rho_{2}\left(|r_{1}-r_{2}|\right)}{\rho_{1}\left(r_{1}\right)}f_{M}\left(\boldsymbol{p}_{2}\right)\right]f^{(1)}\nonumber \\
=-\int\int d\boldsymbol{r}_{2}d\boldsymbol{p}_{2}L_{12}^{QC}\Delta f^{(2)}\\
\left(\frac{\partial}{\partial t}+L_{1}+L_{2}+\bar{L}_{1}^{Q}+\bar{L}_{2}^{Q}\right)f^{(2)}=0,
\end{gather}
where now the closed $n=2$ equation describes the evolution of two
particles under the influence of the potential of mean force. Classically
the Boltzmann equation can be obtained here by taking the so-called
Boltzmann-Grad limit \citep{Grad1958}, with the associated approximation
of molecular chaos (scattering particle initially uncorrelated), this
procedure remains unmodified in the classical MFKT \citep{Baalrud2019}.
The derivation of the BUU equation rests on more uncertain ground.
Originally proposed by Uehling and Uhlenbeck \citep{Uehling1933},
a more rigorous definition was attempted in terms of the Wigner function
in references \citep{Imam-Rahajoe1967,Hoffman1965,Ross1954}. However,
the situation remains unsettled, with the most rigorous derivations
of the BUU equation being obtained in the density-matrix formalism
by Snider \citep{Snider1995} and Boercker and Duffty \citep{Boercker1979}.
The result ultimately is indeed the BUU equation written in terms
of a scattering T-matrix that plays an analogous role to the differential
cross section. There is a caveat: the scattering problem must be solved
with the intermediate states of any perturbation expansion respecting
exchange symmetry. We argue that in our formalism this is accounted
for within the potential itself: the mean-force potential contains
the statistical information regarding spectator particles that ``dress''
the interaction. We further argue that steps in the past derivations
of the BUU equation are not affected negatively by the presence of
the mean-force potential, and can be carried over to our case. The
closure approximation and the molecular chaos assumption still restrict
the range of validity of the theory by limiting the extent to which
correlations are included, which restricts the theory to moderate
values of $\Gamma$, as is also true in the classical case \citep{Baalrud2019}.

The resulting collision integral is that of Uehling and Uhlenbeck
\begin{widetext}
\begin{equation}
C_{q}^{ss'}=\int d\boldsymbol{v}'d\Omega\frac{d\sigma}{d\Omega}u\left[\hat{f_{s}}\hat{f_{s'}}\left(1+\theta_{s}f_{s}\right)\left(1+\theta_{s'}f_{s'}\right)-f_{s}f_{s'}\left(1+\theta_{s}\hat{f_{s}}\right)\left(1+\theta_{s}\hat{f_{s'}}\right)\right],\label{eq:collisionoperator}
\end{equation}
where the ``hatted'' quantities $\hat{f_{s}}$ are evaluated at
the post-collision velocity $\hat{\boldsymbol{v}}=\boldsymbol{v}+\Delta\boldsymbol{v}$
and $\theta_{s}=(\pm1/g_{s})(h/m_{s})^{3}$ where $g_{s}$ is an integer
accounting for particle statistics, the $+$ sign corresponds with
bosons and the $-$ sign with fermions. The difference between this
and the established BUU collision operator is that in the calculation
of the differential cross section $d\sigma/d\Omega$ the scattering
potential is now the potential of mean force. This can be obtained
through solution of the radial Schr\"{o}dinger equation
\begin{equation}
\left\{ \frac{d^{2}}{dr^{2}}+\left[k^{2}-\frac{l(l+1)}{r^{2}}-\frac{2m}{\hbar^{2}}W_{ss'}^{(2)}(r)\right]\right\} R_{l}=0
\end{equation}
via a partial wave expansion, where $W_{ss'}^{(2)}=-k_{B}T\ln g_{ss'}(|r_{1}-r_{2}|)$
is the $n=2$ limit of equation (\ref{eq:PMFcalc}) in terms of the
pair correlation function $g_{ss'}(|r_{1}-r_{2}|)\equiv g_{ss'}^{(2)}(\boldsymbol{r}_{1},\boldsymbol{r}_{2})$.
Calculation of the differential cross section can be performed through
standard quantum scattering theory \citep{Landau1965}.
\end{widetext}

Here we summarize the approximations that have been made to obtain
this collision integral. First, we made the closure approximation
$\Delta f^{(3)}/f_{0}^{(3)}\ll1$, meaning that the triplet distribution
function of a given species is close to its equilibrium limit. Furthermore,
we made the quasi-classical assumption that the product of the degeneracy
and coupling parameters was not overly large. For electron-ion interactions,
this can be justified by the smallness of the electron-ion mass ratio.
Finally, we assume Markovian collisions and molecular chaos, the latter
of which again is justified for the case of electron-ion collisions
(whereas for electron-electron collisions one cannot assume $f^{(2)}(1,2)=f^{(1)}(1)f^{(1)}(2)$
without violating exchange symmetry). These approximations make for
a model that is well-suited for the electron-ion collisions in the
WDM regime.

\subsection{The Potential of Mean Force}

The interpretation of this result is that particles in a strongly
coupled plasma do not collide in isolation. Instead of fully treating
the dynamics of an $n>2$ body collision, the theory shows that the
dynamics are approximated by an effective 2-body interaction obtained
by canonically averaging over the phase-space of the remaining $N-2$
particles at equilibrium. This is similar to the reasoning that leads
to the use of screening potentials of the Yukawa type, and indeed
the mean-force theory reduces to a Debye screened binary collision
theory in the weakly-coupled classical limit. However, the classical
Debye-Huckel, or quantum Thomas-Fermi, potentials do not account for
correlation effects. Correlations are included in our theory through
the presence of the correlation function $\rho_{n}$ in equation (\ref{pomf}).
Specifically, for $2-$body collisions in a uniform plasma, the potential
of mean force can be determined from the correlation function via
equation (\ref{eq:PMFcalc}). The function $g(r)$ is an equilibrium
quantity and accurate methods have already been devised to model this
under a wide range of conditions (see e.g. \citep{Hansen2006} for
the classical limit and \citep{Starrett2013} for a quantum treatment).
Thus, the out-of-equilibrium dynamics have been removed from the problem
of calculating the structure. All that is needed is a method of obtaining
$g(r)$ at equilibrium. This can be provided in principle by a variety
of possible means.

\begin{figure}[b]
\includegraphics[width=8.6cm]{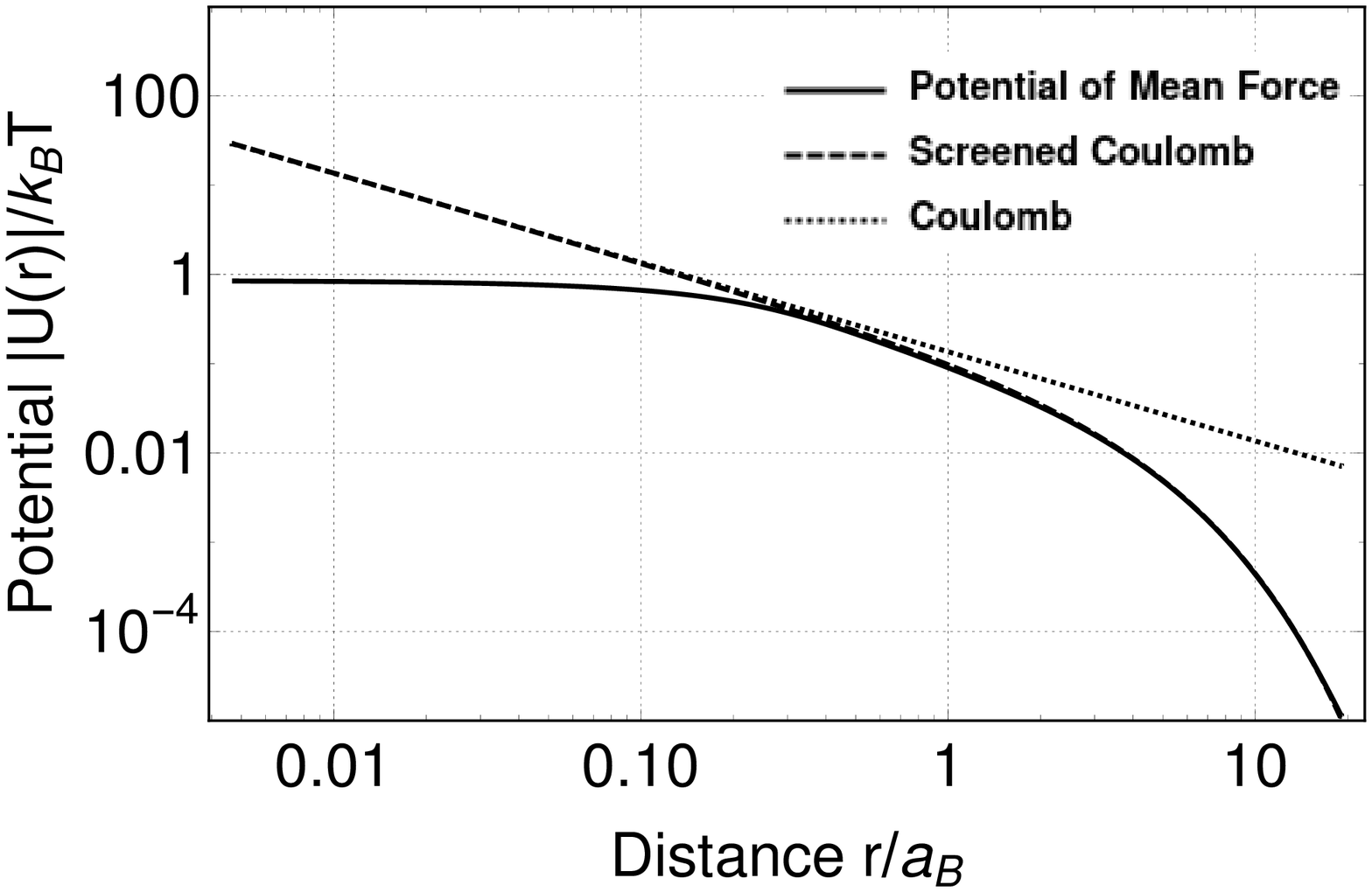}\\
\includegraphics[width=8.6cm]{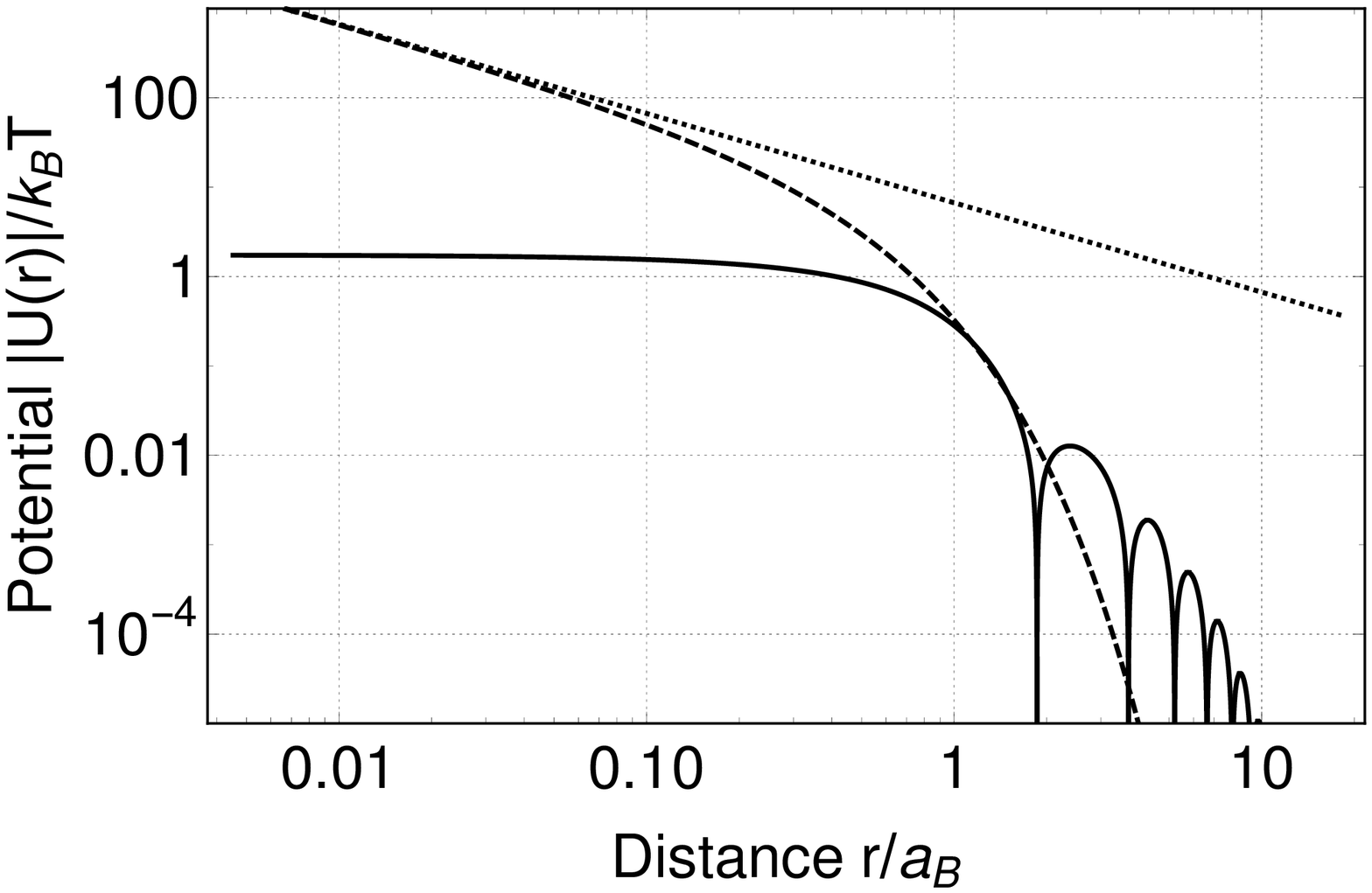}\caption{\label{fig:gii_Vie} Example electron-deuteron scattering potentials
in classical, weakly coupled regime $T=300{\rm eV}$ and $n=5.3\times10^{23}{\rm cm}^{-3}$,
for which $\Theta=13$ and $\Gamma_{e}=\Gamma_{i}=0.043$ (top) and
degenerate, moderately-coupled regime with $T=3{\rm eV}$, $n=4.5\times10^{23}$,
for which $\Theta=0.14$, $\Gamma_{i}=4.1$ and $\Gamma_{e}=1.3$
(bottom). Here, $a_{B}$ is the Bohr radius. Solid line: potential
of mean force; dashed line: screened Coulomb potential with screening
length from equation (\ref{eq:rsc}); dotted line: bare Coulomb potential.}
\end{figure}

For use in populating large tables of transport coefficients over
large parameter spaces, practicality demands fast computation of $g(r)$.
One such method is the hypernetted-chain approximation (HNC) which
has proven successful in the classical case \citep{Baalrud2013,Baalrud2014,Shaffer2019},
and a more sophisticated method applicable to WDM is a coupled average-atom
two-component-plasma quantum HNC method that accounts for electron
degeneracy and partial ionization \citep{Daligault2016,Starrett2012}.
Such methods can be substantially faster than full dynamical calculations
such as molecular dynamics. However, more accurate data can be obtained
from detailed simulations, or even from experiments, in order to verify
the pair distributions used and the results obtained. In figure \ref{fig:gii_Vie}
we show example electron-ion scattering potentials for warm dense
deuterium at conditions that lie in the weakly coupled classical and
moderately coupled degenerate regimes; see regions (a) and (d) of
figure \ref{fig:parregimes}. Potentials shown are the bare Coulomb
potential, which pertains to the original BUU equation, the screened
Coulomb potential
\begin{equation}
U_{{\rm sc}}(r)=\frac{\phi(r)}{k_{{\rm B}}T}{\rm e}^{-r/\lambda_{{\rm sc}}}\label{eq:Uscreen}
\end{equation}
 with degeneracy-dependent screening length
\begin{equation}
\lambda_{{\rm sc}}^{2}=\lambda_{{\rm D}}^{2}\sqrt{\frac{{\rm Li}_{3/2}\left[\exp\left(-\mu/k_{{\rm B}}T\right)\right]}{{\rm Li}_{1/2}\left[\exp\left(-\mu/k_{{\rm B}}T\right)\right]}}\label{eq:rsc}
\end{equation}
with $\lambda_{{\rm D}}$ being the Debye length, and the potential
of mean force derived from pair correlations obtained from a coupled
average-atom two-component-plasma model as described in \citep{Starrett2012,Starrett2013}.
Such models provide fast and accurate computation without the need
for intensive ab-initio simulation. The figure demonstrates the convergence
of the potential of mean force with a screened Coulomb potential in
the weakly-coupled limit, and the importance of correlations in the
calculation of the potential in the region of moderate coupling.

\section{Transport Rates\label{sec:Relaxation}}

\subsection{General Formalism}

Temperature relaxation is a transport process of practical importance
to WDM. It acts as a probe of electron-ion collisions in this environment,
and the relaxation time $\tau$ serves as an important indicator of
the underlying physics. WDM is often created in a state that is far
from equilibrium. Due to the electron-ion mass ratio, electrons and
ions quickly relax into separate populations at two different temperatures.
Electron-ion scattering then more slowly equilibrates the two temperatures.
The rate $\tau$ is additionally an experimentally accessible parameter.
Recent experiments have begun measuring this relaxation in the WDM
regime \citep{Cho2016,Zaghoo2019}, and are showing interesting effects
due to both degeneracy and Coulomb coupling. Electron-ion relaxation
has been difficult to address computationally as a consequence of
the mass ratio along with degeneracy and correlation. However, recent
advances have been made to finally make reliable predictions of relaxation
rates using wave packet molecular dynamics \citep{Ma2019} and mixed
classical-quantum molecular dynamics \citep{Simoni2019,Daligault2018a}.
These methods are opening the door for unprecedented validation of
analytical and numerical models. Theoretical prediction of relaxation
in a plasma is typified by the theory of Landau and Spitzer. The Landau-Spitzer
(LS) theory depends on the Coulomb logarithm $\ln\Lambda_{ei}$, which
vanishes in dense plasmas when the screening length becomes smaller
than the Wigner-Seitz radius $a$. Current models of temperature relaxation
include the Fermi Golden Rule (FGR) method \citep{Hazak2001,DharmaWardana1998},
a standard plasma approach using the quantum Landau equation \citep{Daligault2016a},
a method based on a quantum field theory for plasmas \citep{Brown2005,Brown2007,Singleton2007},
and an approach that incorporates correlation effects in a normal
mode kinetic calculation through including local field corrections
(LFCs) in the dielectric response \citep{Daligault2009}, and the
quantum Lenard-Balescu equation \citep{Benedict2012,Scullard2018}.
Finally, a new method provides the relaxation rate in terms of the
electron-ion collision cross section \citep{Daligault2019} and allows
for both degeneracy and correlation in a large area of parameter space.
This approach treats the ions as classical particles obeying a Langevin-like
equation while the populations of the electron states obey a master
equation, with relaxation rates determined from the frictional forces
felt by the ions.

Momentum relaxation is another important process in dense plasmas.
A velocity drift between electron and ion populations gives rise to
a current, and electron-ion collisional friction contributes to the
conductivity. While it should be noted that electron-electron collisions
also contribute in calculations of the conductivity, in certain regimes
the electron-ion collisions are dominant. LS theory again provides
the conductivity for a classical plasma, but neglects both correlation
and quantum effects. To correct for this, Lee and More introduced
a transport model attempting to extend LS theory to regimes with strong
coupling and degeneracy \citep{Lee1984}. The quantum LFP equation
has been applied to the calculation of momentum relaxation as well
\citep{Daligault2016a}. However, these methods do not accurately
account for strong collisions. The effects of strong collisions with
diffraction on temperature relaxation have been addressed by the so-called
T-matrix approach, but without incorporating degeneracy \citep{Gericke2002}.
Degeneracy has been addressed through a variety of methods. The quantum
Lenard-Balescu equation has been applied to electrical conductivity
\citep{Lampe1968}, but again is limited to weak coupling. A model
for conductivity in a degenerate and strongly-coupled plasma was obtained
via the Kubo-Greenwood formalism combined with an average-atom two-component
plasma model, where the degeneracy is included through its influence
on the electron-ion potential \citep{Starrett2018,Gill2019}.

A binary mixture of two species $s$ and $s'$ out of equilibrium
will relax towards equilibrium through $s-s$, $s-s'$ and $s'-s'$
collisions, which are modeled by the collision operator (\ref{eq:collisionoperator}).
Specifically, taking velocity moments of the kinetic equation produces
a hierarchy of fluid equations which can be closed through a perturbative
expansion about equilibrium. Transport coefficients are determined
from moments of the collision operator. The relevant moments can be
written
\begin{equation}
\left\langle \chi_{l}\right\rangle ^{s-s'}=\int d\boldsymbol{v}\chi\left(\boldsymbol{v}\right)C_{qB}^{s-s'},
\end{equation}
where $\chi(\boldsymbol{v})$ is some polynomial function of the velocity.
To simplify we utilize the following properties: $d\Omega\frac{d\sigma}{d\Omega}$
is invariant under reversal of the collision, i.e. $\left(\boldsymbol{\boldsymbol{v}},\boldsymbol{v}'\right)\leftrightarrow\left(\hat{\boldsymbol{v}},\hat{\boldsymbol{v}}'\right)$
where $\boldsymbol{\boldsymbol{v}}\thinspace{\rm and\thinspace}\boldsymbol{v}'$
are the pre-collision velocities of particles one and two respectively
and the ``hat'' $\hat{}$ indicates a post-collision quantity, and
the phase-space volume element is invariant, i.e. $\int d\boldsymbol{v}d\boldsymbol{v}'=\int d\hat{\boldsymbol{v}}d\hat{\boldsymbol{v}'}$.
We thus obtain
\begin{gather}
\left\langle \chi_{l}\right\rangle ^{s-s'}=\int d\boldsymbol{v}\int d\Omega\frac{d\sigma}{d\Omega}u\int d\boldsymbol{v}'\left[\chi\left(\hat{\boldsymbol{v}}\right)-\chi\left(\boldsymbol{v}\right)\right]\nonumber \\
\times f_{s}f_{s'}\left(1+\theta_{s}\hat{f_{s}}\right)\left(1+\theta_{s}\hat{f_{s'}}\right).\label{eq:chimoments}
\end{gather}
Relevant $\chi\left(v\right)$ are
\begin{widetext}
\begin{equation}
\chi\left(\boldsymbol{v}\right)=\begin{cases}
1\thinspace\thinspace\thinspace\thinspace\thinspace\thinspace\thinspace\thinspace\thinspace\rightarrow & \left[\chi\left(\hat{\boldsymbol{v}}\right)-\chi\left(\boldsymbol{v}\right)\right]=0\\
m_{s}\boldsymbol{v}\thinspace\rightarrow & \left[\chi\left(\hat{\boldsymbol{v}}\right)-\chi\left(\boldsymbol{v}\right)\right]=m_{s}\boldsymbol{\Delta}v\\
m_{s}v^{2}\rightarrow & \left[\chi\left(\hat{\boldsymbol{v}}\right)-\chi\left(\boldsymbol{v}\right)\right]=m_{s}\Delta v^{2}
\end{cases}
\end{equation}
where $\Delta\boldsymbol{v}=\hat{\boldsymbol{v}}-\boldsymbol{v}$.
Substituting variables $\boldsymbol{v}=\boldsymbol{v}'+\boldsymbol{u}$
and the collision kinematics which determine
\begin{equation}
m_{s}\Delta\boldsymbol{v}=m_{ss'}\Delta\boldsymbol{u},
\end{equation}
\begin{equation}
\Delta\boldsymbol{u}\cdot\Delta\boldsymbol{u}=-2\boldsymbol{u}\cdot\Delta\boldsymbol{u}
\end{equation}
and
\begin{equation}
\left(2\boldsymbol{v}\cdot\Delta\boldsymbol{v}+\Delta\boldsymbol{v}^{2}\right)=\frac{m_{ss'}}{m_{s}}\Delta\boldsymbol{u}\cdot\left(\boldsymbol{v}'+\frac{m_{ss'}}{m_{s}}\boldsymbol{u}\right),
\end{equation}
(see \citep{Baalrud2012}) shows that
\begin{equation}
\chi\left(\boldsymbol{v}\right)=\begin{cases}
1\thinspace\thinspace\thinspace\thinspace\thinspace\thinspace\thinspace\thinspace\thinspace\rightarrow & \left[\chi\left(\hat{\boldsymbol{v}}\right)-\chi\left(\boldsymbol{v}\right)\right]=0\\
m_{s}\boldsymbol{v}\thinspace\rightarrow & \left[\chi\left(\hat{\boldsymbol{v}}\right)-\chi\left(\boldsymbol{v}\right)\right]=m_{ss'}\Delta\boldsymbol{u}\\
m_{s}v^{2}\rightarrow & \left[\chi\left(\hat{\boldsymbol{v}}\right)-\chi\left(\boldsymbol{v}\right)\right]=m_{ss'}\left(\boldsymbol{v'}-\boldsymbol{V}_{s}+\frac{m_{ss'}}{m_{s'}}\boldsymbol{u}\right)\cdot\Delta\boldsymbol{u}
\end{cases}
\end{equation}
where 
\begin{equation}
\Delta\boldsymbol{u}=u\left({\rm sin}\theta{\rm cos}\phi\hat{\boldsymbol{x}}+{\rm sin}\theta{\rm sin}\phi\hat{\boldsymbol{y}}-2{\rm sin}^{2}\frac{\theta}{2}\hat{\boldsymbol{u}}\right).
\end{equation}
\end{widetext}

The preceding discussion and the collision operator (\ref{eq:collisionoperator})
are in principle applicable to transport in any semi-classical system.
As it pertains to WDM, ion-ion scattering is contained within this
formalism as ion dynamics are classical and electron degeneracy effects
enter only via the potential of mean force. Application of the theory
to ion-ion scattering was validated in \citep{Daligault2016}. The
case of the electron-electron terms requires further work due to the
subtleties discussed at the end of section \ref{sec:theory}. The
model at the level to which we have developed it has immediate applicability
to the case of electron-ion scattering.

\subsection{The Relaxation Problem}

Presently, we restrict our analysis to the class of problems in which
electrons and ions in the plasma are in respective equilibrium with
themselves with different fluid quantities $T_{e},\thinspace T_{i},\thinspace\boldsymbol{V}_{e}\thinspace{\rm and}\thinspace\boldsymbol{V}_{i}$,
respectively. In such a system, the electron and ion fluid variables
will equilibrate on a timescale long compared to the respective electron-electron
and ion-ion collision times. The ions have a classical Maxwellian
velocity distribution
\begin{equation}
f_{i}\left(\boldsymbol{v}'\right)=\frac{n_{i}}{v_{Ti}^{3}}\frac{e^{-\left(\boldsymbol{v}'-\boldsymbol{V}_{i}\right)^{2}/v_{Ti}^{2}}}{\pi^{3/2}}
\end{equation}
and the electrons have an appropriately normalized Fermi-Dirac velocity
distribution
\begin{equation}
f_{e}\left(\boldsymbol{v}\right)=n_{e}\left[v_{Te}^{3}\left(-\pi^{3/2}\text{Li}_{\frac{3}{2}}(-\xi)\right)\left(1+\frac{{\rm e}^{\left(\boldsymbol{v}-\boldsymbol{V}_{e}\right)^{2}/v_{Te}^{2}}}{\xi}\right)\right]^{-1}
\end{equation}
where $v_{Ts}=\sqrt{2k_{B}T_{s}/m_{s}}$ and $\xi=\exp\left(\mu/k_{B}T\right)$,
the ion velocity is $\boldsymbol{v'}$ and electron velocity is $\boldsymbol{v}$.
We can write
\begin{gather}
f_{e}f_{i}\left(1+\theta_{e}\hat{f_{e}}\right)=\frac{n_{i}}{v_{Ti}^{3}}\frac{e^{-\left(\boldsymbol{v}'-\boldsymbol{V}_{i}\right)^{2}/v_{Ti}^{2}}}{\pi^{3/2}}n_{e}\nonumber \\
\times\left[v_{Te}^{3}\left(-\pi^{3/2}\text{Li}_{\frac{3}{2}}(-\xi)\right)\left(1+\frac{{\rm e}^{\left(\boldsymbol{v}-\boldsymbol{V}_{e}\right)^{2}/v_{Te}^{2}}}{\xi}\right)\right]^{-1}\nonumber \\
\times\left[1-\left(1+\frac{{\rm e}^{\left(\boldsymbol{v}+\Delta\boldsymbol{v}-\boldsymbol{V}_{e}\right)^{2}/v_{Te}^{2}}}{\xi}\right)^{-1}\right],\label{eq:integrand1}
\end{gather}
from which the relation of the factor $\left(1+\theta_{e}\hat{f_{e}}\right)$
to Pauli blocking can be seen in terms of the Fermi-Dirac occupation
number: the contribution to the collision integral from collisions
to or from occupied states is zero. This simplification occurs from
the combination of $\theta_{e}$ with the Fermi Dirac distribution
and the relation (\ref{eq:xi_Theta_relation}). 

Electron-ion temperature and momentum relaxation rates depend on the
energy exchange density $Q^{s-s'}$ and friction force density $\boldsymbol{R}^{s-s'}$,
respectively. These can in turn be written in terms of the moments
(\ref{eq:chimoments}), assuming a uniform plasma, as 
\begin{equation}
Q^{ei}=\left\langle \frac{1}{2}m_{e}\left(\boldsymbol{v}-\boldsymbol{V}_{e}\right)^{2}\right\rangle ^{e-i}=\frac{3n_{e}}{2}\frac{dT_{e}}{dt}\label{eq:Qei}
\end{equation}
(where in the last equality we have taken $\boldsymbol{V}_{e}=\boldsymbol{V}_{i}=0)$
and 
\begin{equation}
\boldsymbol{R}^{ei}=\left\langle m_{e}\boldsymbol{v}\right\rangle ^{e-i}=m_{e}d\left\langle \boldsymbol{v}\right\rangle /dt=m_{e}\frac{d\boldsymbol{V}_{e}}{dt}\label{eq:Rei}
\end{equation}
which, in the respective limits of $\Delta T\ll T$ and $\Delta V\ll V$
yield simple relaxation rates $dT_{e}/dt=\nu_{ei}^{(\epsilon)}\Delta T$
and $d\boldsymbol{V}_{e}/dt=\nu_{ei}^{(p)}\Delta\boldsymbol{V}$.

The integration over the ion velocity can be simplified significantly
in the limit that the ion velocities are much smaller than the electron
velocities: $m_{e}T_{i}\ll m_{i}T_{e}$, which is true when temperature
differences are not extreme, which coincides with our expansion about
the equilibrium state. By expanding equation (\ref{eq:integrand1})
in the limit that the electron distribution is approximately constant
over the range of accessible ion velocities, the integral over the
ion velocities can be carried out analytically. The evaluation of
this integral differs for the calculation of $Q^{ei}$ versus $\boldsymbol{R}^{ei}$.
Therefore we examine each case separately.

\subsubsection{Temperature Relaxation}

The energy-exchange density (\ref{eq:Qei}) in this case becomes 
\begin{gather*}
Q^{ei}=\int d\boldsymbol{u}\int d\Omega\frac{d\sigma}{d\Omega}u\boldsymbol{\Delta u}\\
\cdot\int d\boldsymbol{v}'m_{ei}\left(\boldsymbol{v'}+\frac{m_{ei}}{m_{i}}\boldsymbol{u}\right)f_{i}f_{e}\left(1-|\theta|\hat{f_{e}}\right).
\end{gather*}
Inserting equation (\ref{eq:integrand1}) and applying the expansion
$|\boldsymbol{v}'|\ll|\boldsymbol{u}|$ and assuming zero drift velocities
and $\left|T_{e}-T_{i}\right|\ll T_{e},T_{i}$ we perform the integral
over $\boldsymbol{v}'$ and write
\begin{gather*}
\boldsymbol{\Delta u}\cdot\int d\boldsymbol{v}'m_{ei}\left(\boldsymbol{v'}+\frac{m_{ei}}{m_{i}}\boldsymbol{u}\right)f_{i}f_{e}\left(1-|\theta|\hat{f_{e}}\right)\approx\\
\frac{n_{e}n_{i}\xi e^{-\eta^{2}}\sin^{2}\left(\frac{\theta}{2}\right)}{\pi^{3}v_{Te}\text{Li}_{\frac{3}{2}}\left(-\xi\right)\left(\xi e^{-\eta^{2}}+1\right)^{2}}
\end{gather*}
where $\eta\equiv u/v_{Te}$ and therefore obtain, written to facilitate
comparison with the classical limit,
\begin{equation}
Q^{ei}=-3\frac{m_{e}}{m_{i}}n_{e}\nu_{ei}^{(\epsilon)}(T_{e}-T_{i}).\label{eq:Trelax}
\end{equation}
The collision frequency $\nu_{ei}^{(\epsilon)}$ can be written in
terms of a reference frequency, 
\begin{equation}
\nu_{0}\equiv\frac{4\sqrt{2\pi}n_{e}e^{4}}{3\sqrt{m_{e}}(k_{B}T_{e})^{3/2}}
\end{equation}
 and a generalized Coulomb integral $\Xi_{ei}^{(\epsilon)}$,
\begin{equation}
\nu_{ei}^{(\epsilon)}=\nu_{0}\Xi_{ei}^{(\epsilon)}.
\end{equation}
Effects of degeneracy and strong coupling are contained in the function
which replaces the Coulomb logarithm,
\begin{gather}
\Xi_{ei}^{(\epsilon)}=\frac{1}{2}\int_{0}^{\infty}d\eta G_{1}(\eta)\frac{\sigma_{1}^{(1)}\left(\eta,\Gamma\right)}{\sigma_{0}}\label{eq:xiE1}\\
\equiv\frac{1}{2}\int_{0}^{\infty}d\eta I_{\epsilon}(\eta)\nonumber 
\end{gather}
where
\begin{equation}
\sigma_{1}^{(1)}\left(\eta,\Gamma\right)=4\pi\int_{0}^{\pi}d\theta\sin^{2}\frac{\theta}{2}\sin\theta\frac{d\sigma}{d\Omega}
\end{equation}
is the momentum transfer cross section, written in terms of the phase
shifts $\delta_{l}$ as
\begin{equation}
\sigma_{1}^{(1)}=\frac{4\pi}{\eta^{2}}\sum_{l=0}^{\infty}(l+1)\sin^{2}(\delta_{l+1}-\delta_{l}).\label{eq:momCC1}
\end{equation}
 The function
\begin{equation}
G_{1}(\eta)\equiv\frac{\xi{\rm e}^{-\eta^{2}}\eta^{5}}{\left[-\text{Li}_{\frac{3}{2}}\left(-\xi\right)\right]\left(\xi e^{-\eta^{2}}+1\right)^{2}}
\end{equation}
determines the relative availability of states that contribute to
the scattering. This is plotted in figure \ref{fig:integrands} for
several values of the degeneracy parameter $\Theta$, where it is
shown that in the classical limit scattering is dominated by energy
transfers around the thermal energy, and as degeneracy increases the
envelope of relevant momentum-transfers narrows about the Fermi momentum.
It must be noted that the relaxation rate obtained is identical to
that obtained in equation (71) of reference \citep{Daligault2019}
by very different means, although the PMF does not appear naturally
in this other formulation and must be manually included in the calculations
of the cross sections.

\subsubsection{Momentum Relaxation}

Electron-ion momentum relaxation is a major contributor to electrical
resistivity, diffusion, and is also related to the stopping power
of ions in a plasma. Here, we concentrate on the case that drift velocities
are much slower than the electron thermal speed, with the goal of
obtaining a relaxation rate rather than stopping power. Relaxation
occurs through collisions between electron and ion populations with
different average velocities. The force density (\ref{eq:Rei}) associated
with these collisions is

\begin{equation}
\boldsymbol{R}^{ei}=\int d\boldsymbol{v}\int d\Omega\frac{d\sigma}{d\Omega}u\int d\boldsymbol{v}'m_{ei}\boldsymbol{\Delta u}f_{e}f_{i}\left(1+\theta_{e}\hat{f_{e}}\right).
\end{equation}
Inserting equation (\ref{eq:integrand1}) into the above equation
and applying the expansion $|\boldsymbol{v}'|\ll|\boldsymbol{u}|$
and assuming $\left|T_{e}-T_{i}\right|\ll T_{e},T_{i}$ and $V_{i}\ll v_{Ti}$
and $V_{e}\ll v_{Te}$, the integral over $\boldsymbol{v}'$ can be
performed analytically and results in
\begin{gather*}
\int d\boldsymbol{v}'m_{ei}f_{e}f_{i}\left(1+\theta_{e}\hat{f_{e}}\right)\\
\approx\frac{2m_{e}n_{e}\xi n_{i}e^{-\eta^{2}}\left[\boldsymbol{u}\cdot\boldsymbol{\Delta V}-\xi e^{-\eta^{2}}\left(\boldsymbol{\Delta u}+\boldsymbol{u}\right)\cdot\boldsymbol{\Delta V}\right]}{\pi^{3/2}v_{Te}^{5}\left[-\text{Li}_{\frac{3}{2}}\left(-\xi\right)\right]\left(\xi e^{-\eta^{2}}+1\right){}^{3}}.
\end{gather*}
We follow the classical example and write
\begin{equation}
\boldsymbol{R}^{ei}=-n_{e}m_{e}\nu_{ei}^{(p)}(\boldsymbol{V}_{e}-\boldsymbol{V}_{i})\label{eq:momrelax}
\end{equation}
where the frequency
\begin{equation}
\nu_{ei}^{(p)}=\nu_{0}\Xi_{ei}^{(p)}
\end{equation}
involves a Coulomb integral
\begin{gather}
\Xi_{ei}^{(p)}=\frac{1}{2}\int_{0}^{\infty}d\eta\left[G_{2}(\eta,\xi)\frac{\sigma_{1}^{(1)}\left(\eta,\Gamma\right)}{\sigma_{0}}-G_{3}(\eta,\xi)\frac{\sigma_{2}^{(1)}\left(\eta,\Gamma\right)}{\sigma_{0}}\right]\nonumber \\
=\frac{1}{2}\int_{0}^{\infty}d\eta G_{2}(\eta,\xi)\frac{\sigma_{p}(\eta,\Gamma,\xi)}{\sigma_{0}}\label{eq:xiMOM1}\\
\equiv\frac{1}{2}\int_{0}^{\infty}d\eta I_{p}(\eta,\Gamma,\xi),\nonumber 
\end{gather}
which is different from that involved in the energy-exchange density.
Here, $\sigma_{1}^{(1)}$ was defined previously and
\begin{gather}
\sigma_{2}^{(1)}\left(\eta,\Gamma\right)=4\pi\int_{0}^{\pi}d\theta\sin^{2}\frac{\theta}{2}\sin\theta\frac{d\sigma}{d\Omega}\cos\theta.\label{eq:momCC2a}
\end{gather}
This differs from the classical case in which momentum and temperature
relaxation differ only by a numerical factor and a mass ratio \citep{Baalrud2012}.
The weighting functions 
\begin{gather}
G_{2}(\eta,\xi)=\frac{\xi{\rm e}^{-\eta^{2}}\eta^{5}}{\left[-\text{Li}_{\frac{3}{2}}\left(-\xi\right)\right]\left(\xi e^{-\eta^{2}}+1\right)^{3}},\\
G_{3}(\eta,\xi)=\frac{\xi^{2}{\rm e}^{-2\eta^{2}}\eta^{5}}{\left[-\text{Li}_{\frac{3}{2}}\left(-\xi\right)\right]\left(\xi e^{-\eta^{2}}+1\right)^{3}},
\end{gather}
are shown in figure \ref{fig:integrands} and compared with the statistical
weighting factors in the case of temperature relaxation. The presence
of the differing angular integrals between the energy and momentum
relaxation cases warrants further discussion.

The function (\ref{eq:momCC2a}) behaves differently than a momentum
transfer cross-section, and indeed can take on negative values. Through
the use of the Wigner-3j function, $\sigma_{2}^{(1)}$ can be expanded
in the phase shifts (see appendix \ref{sec:shifts_appendix}) as
\begin{widetext}
\begin{gather}
\sigma_{2}^{(1)}=\frac{4\pi}{\eta^{2}}\sum_{l=0}^{\infty}\frac{\sin\delta_{l}}{4l(l+1)-3}\nonumber \\
\times\left\{ (l+1)(2l-1)\left[(l+2)\sin(\delta_{l}-2\delta_{l+2})-(2l+3)\sin(\text{\ensuremath{\delta_{l}}}-2\text{\ensuremath{\delta_{l+1}}})\right]-l^{2}(2l+3)\sin\delta_{l}\right\} .\label{eq:momCC2b}
\end{gather}
While it is tempting to interpret the quantity $\sigma_{2}^{(1)}$
as a cross-section, $\sigma_{2}^{(1)}$ can become negative, and as
will be shown in the following section, it is only in the combination
$\sigma_{p}=\sigma_{1}^{(1)}\left(\eta,\Gamma\right)-\xi_{e}{\rm e}^{-\eta^{2}}\sigma_{2}^{(1)}\left(\eta,\Gamma\right)$
that this interpretation is justified. We thus refer to $\sigma_{p}$
as an effective transport cross section. We note that this second
term arises due to degeneracy, and has no analog in the classical
relaxation problem.
\begin{figure}[b]
\includegraphics[width=8.6cm]{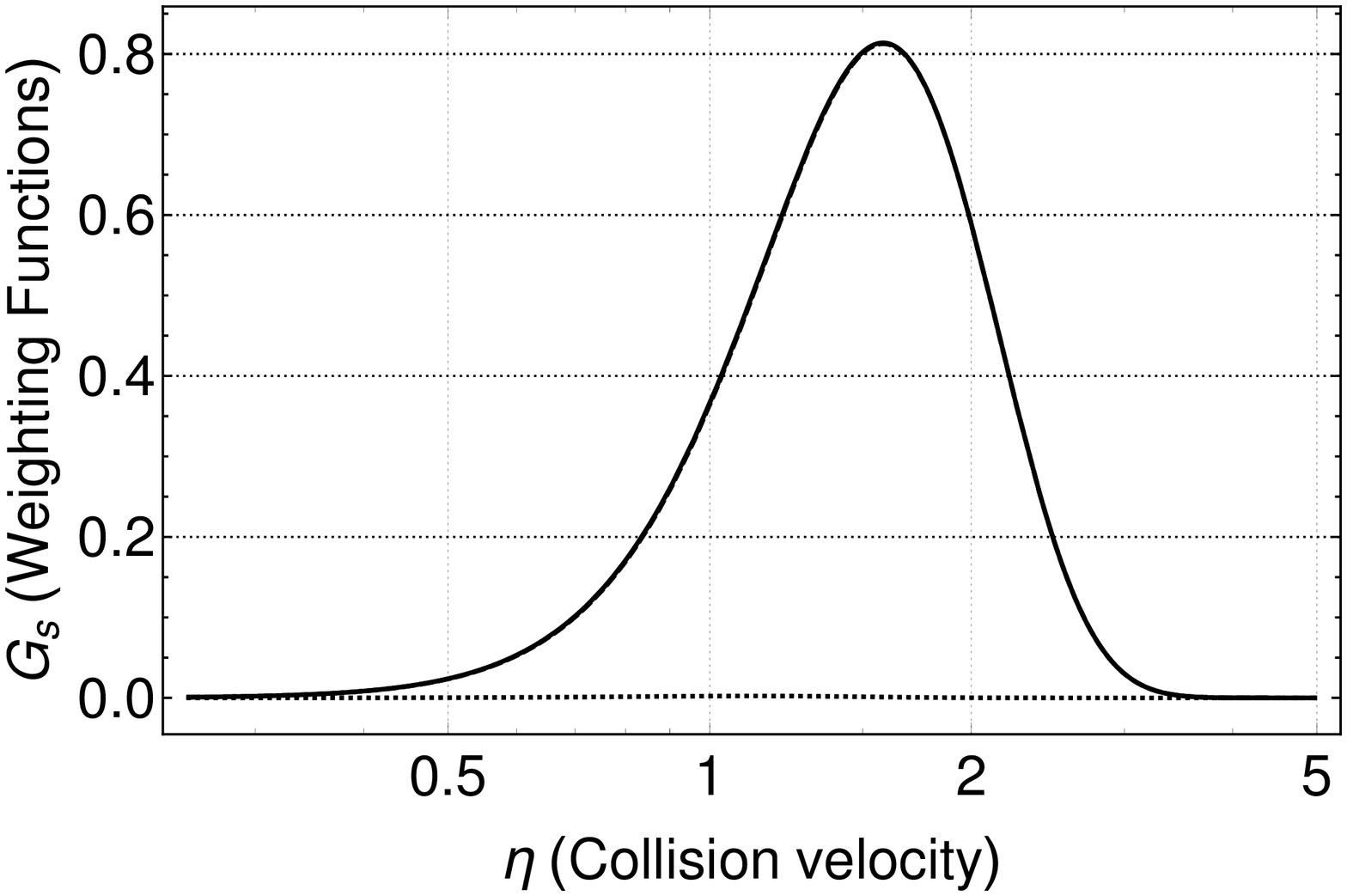}\\
\includegraphics[width=8.6cm]{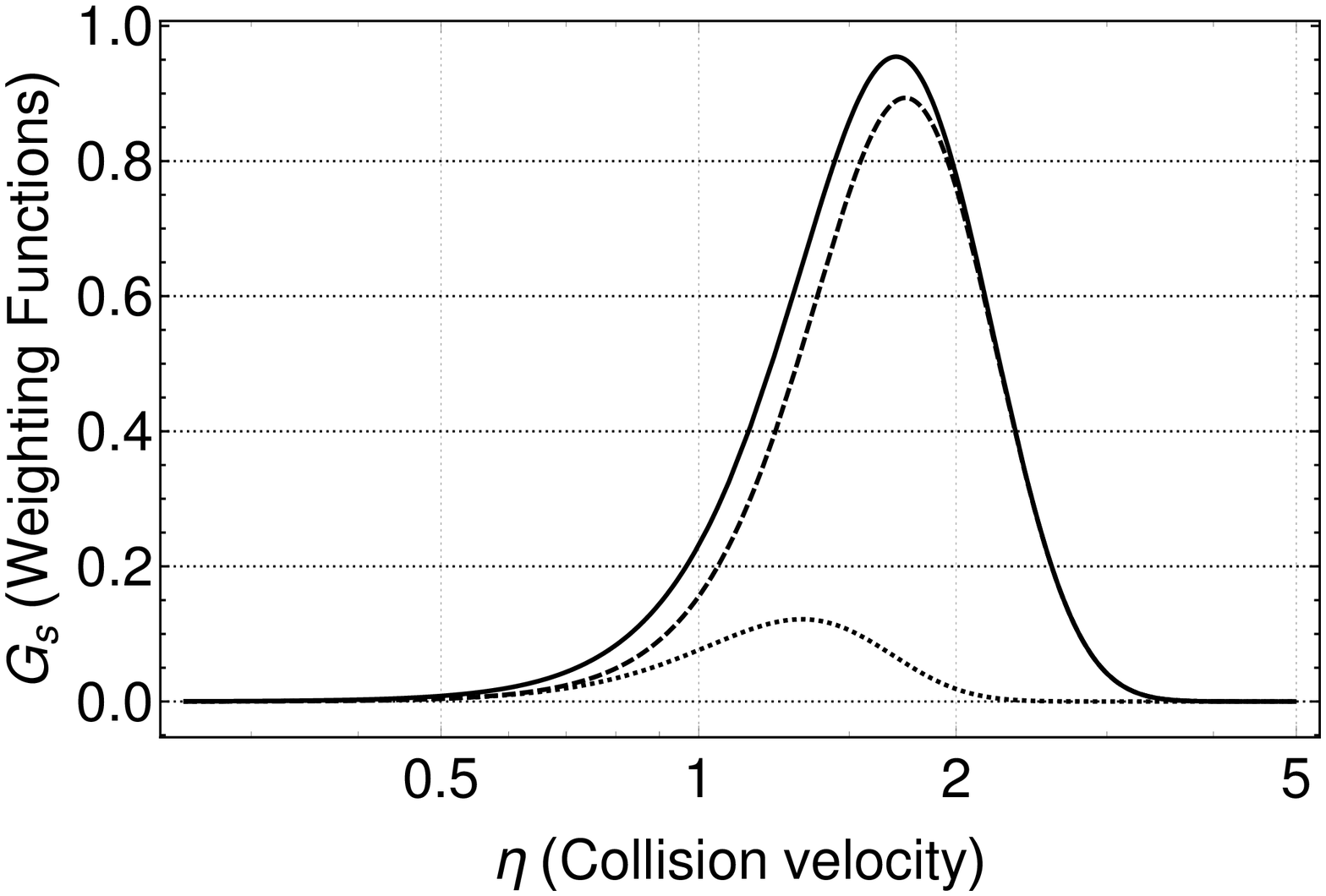}\\
\includegraphics[width=8.6cm]{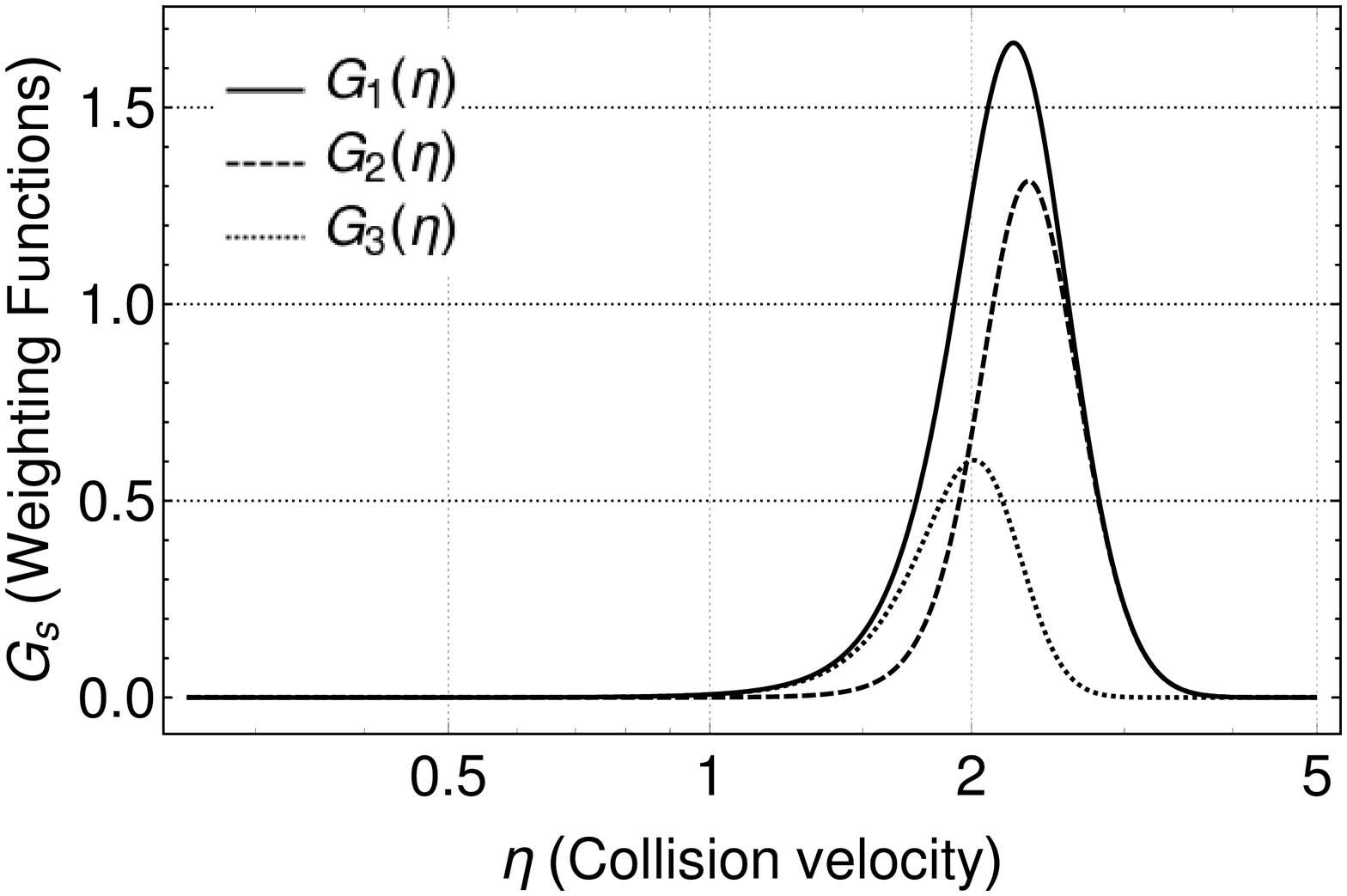}

\caption{\label{fig:integrands} Statistical contributions to the integrands
for temperature and momentum relaxation, $G_{1}$ (solid), $G_{2}$
(dashed) and $G_{3}$ (dotted), for three conditions: $\Theta=12.6$
and $\xi=0.017$ (top, weak degeneracy), $\Theta=0.85$ and $\xi=1.33$
(middle, moderate degeneracy), $\Theta=0.14$ and $\xi=1135$ (bottom,
strong degeneracy). The relevant collision velocities for both momentum
and temperature relaxation become narrowly centered around the Fermi
velocity at strong degeneracy. The relative importance of the two
different functions that contribute to momentum relaxation can be
seen to depend on the degeneracy.}
\end{figure}
\end{widetext}

\begin{figure*}
\includegraphics[width=8.6cm]{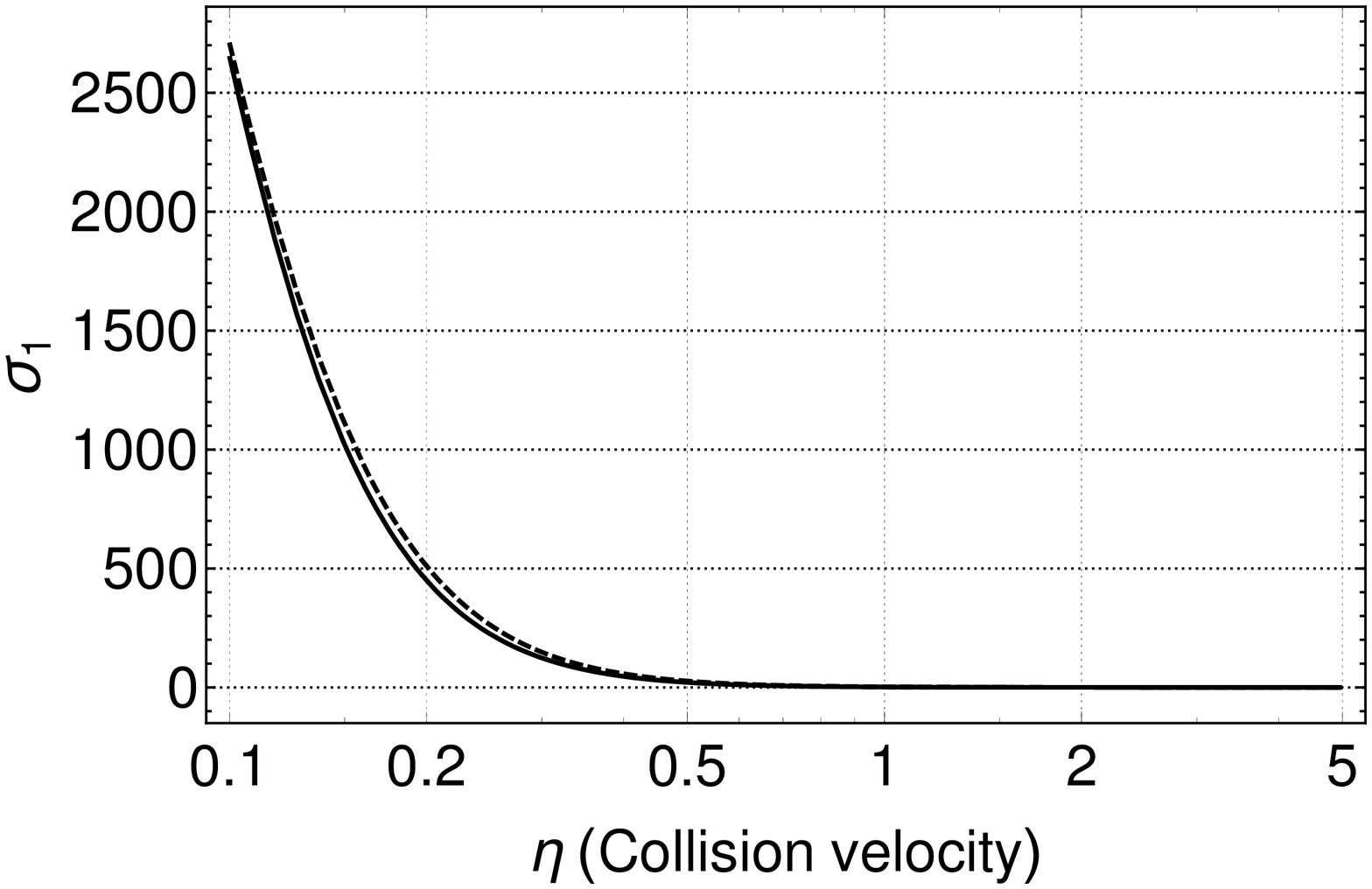}\includegraphics[width=8.6cm]{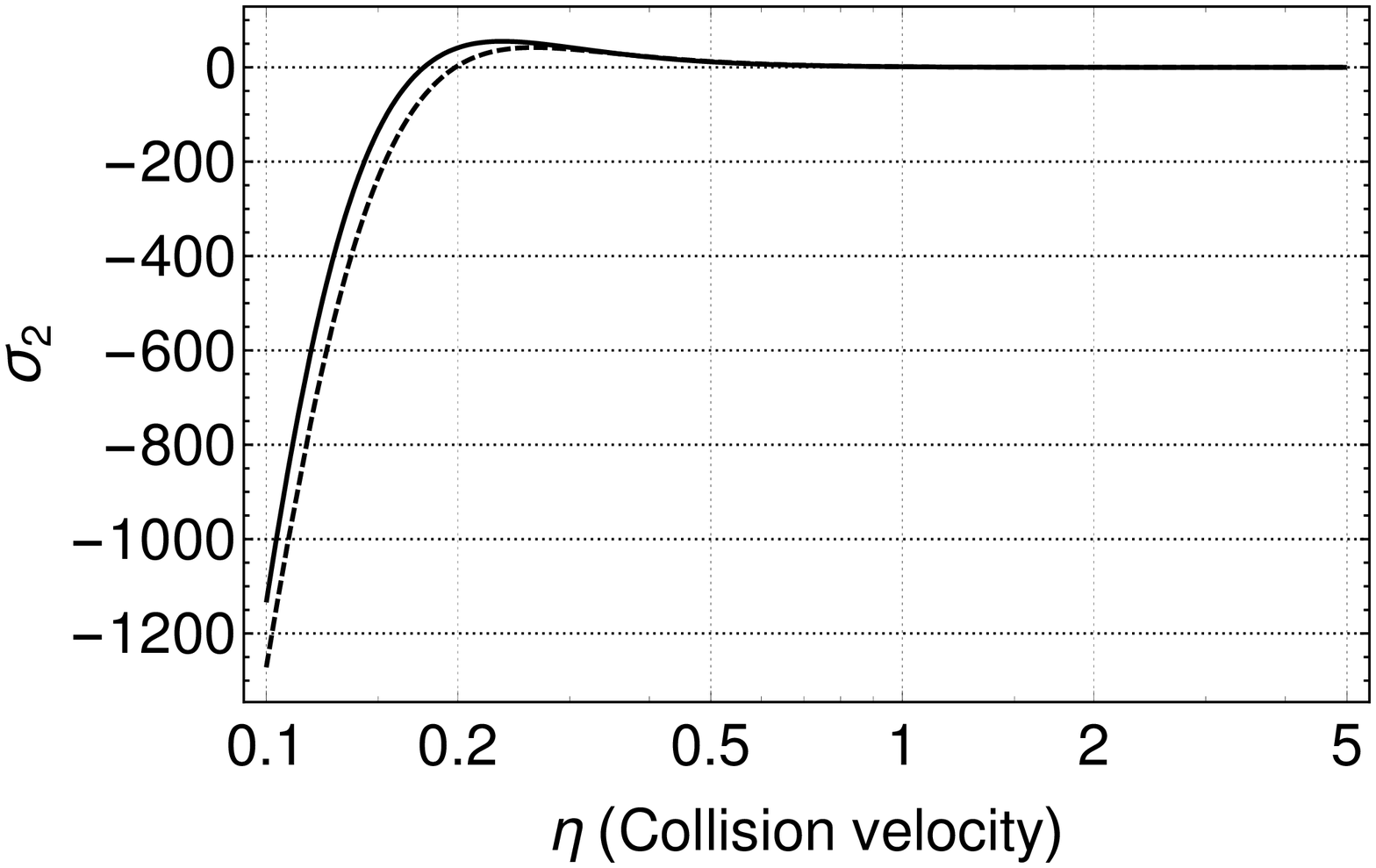}\\
\includegraphics[width=8.6cm]{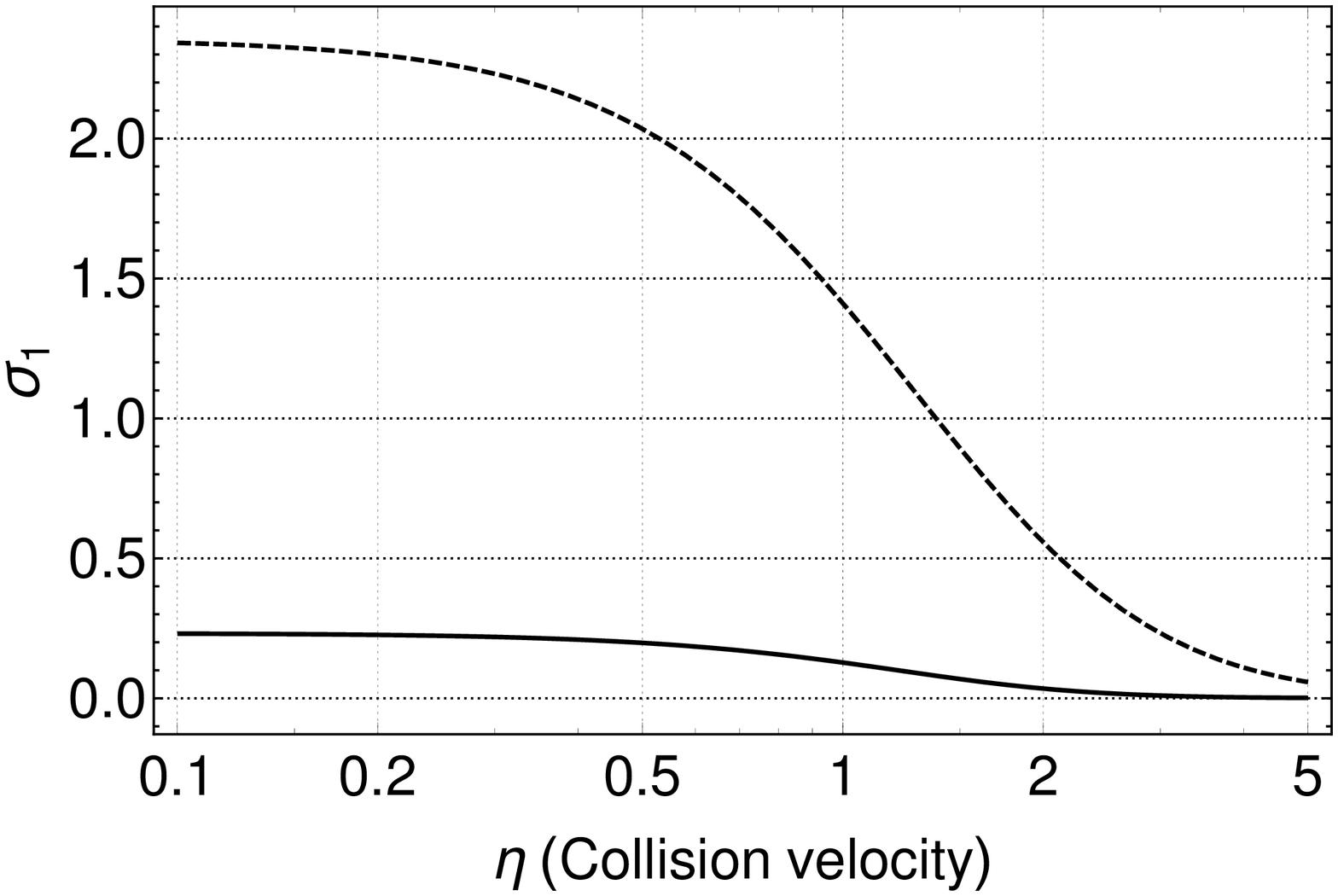}\includegraphics[width=8.6cm]{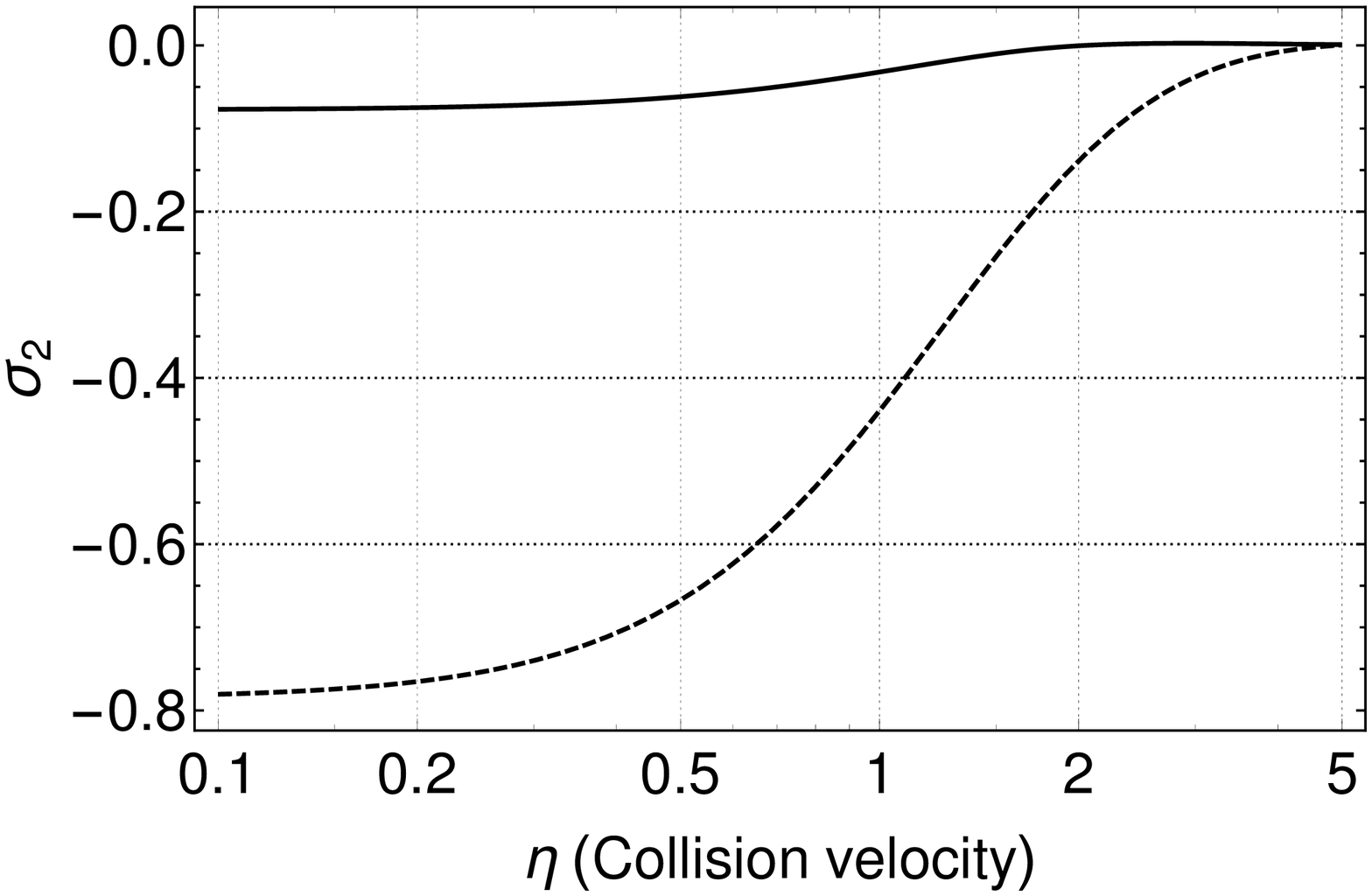}\caption{\label{fig:momCC} Top: Functions $\sigma_{1}^{(1)}$ (left) and $\sigma_{2}^{(1)}$
(right) calculated using the PMF (solid) and a screened Coulomb potential
equation (\ref{eq:Uscreen}) (dashed) in the classical, weakly-coupled
limit with $T=200\thinspace{\rm eV}$ and $n=0.774\thinspace{\rm g}\cdot{\rm cm}^{-3}$.
Bottom: the same, but in the degenerate, moderately-coupled case with
$T=3\thinspace{\rm eV}$ at the same density. This shows the different
behaviors of the effective transport cross sections, and the breakdown
of the screened Coulomb potential with the appearance of coupling
effects at low temperature.}
\end{figure*}

\begin{figure}[b]
\includegraphics[width=8.6cm]{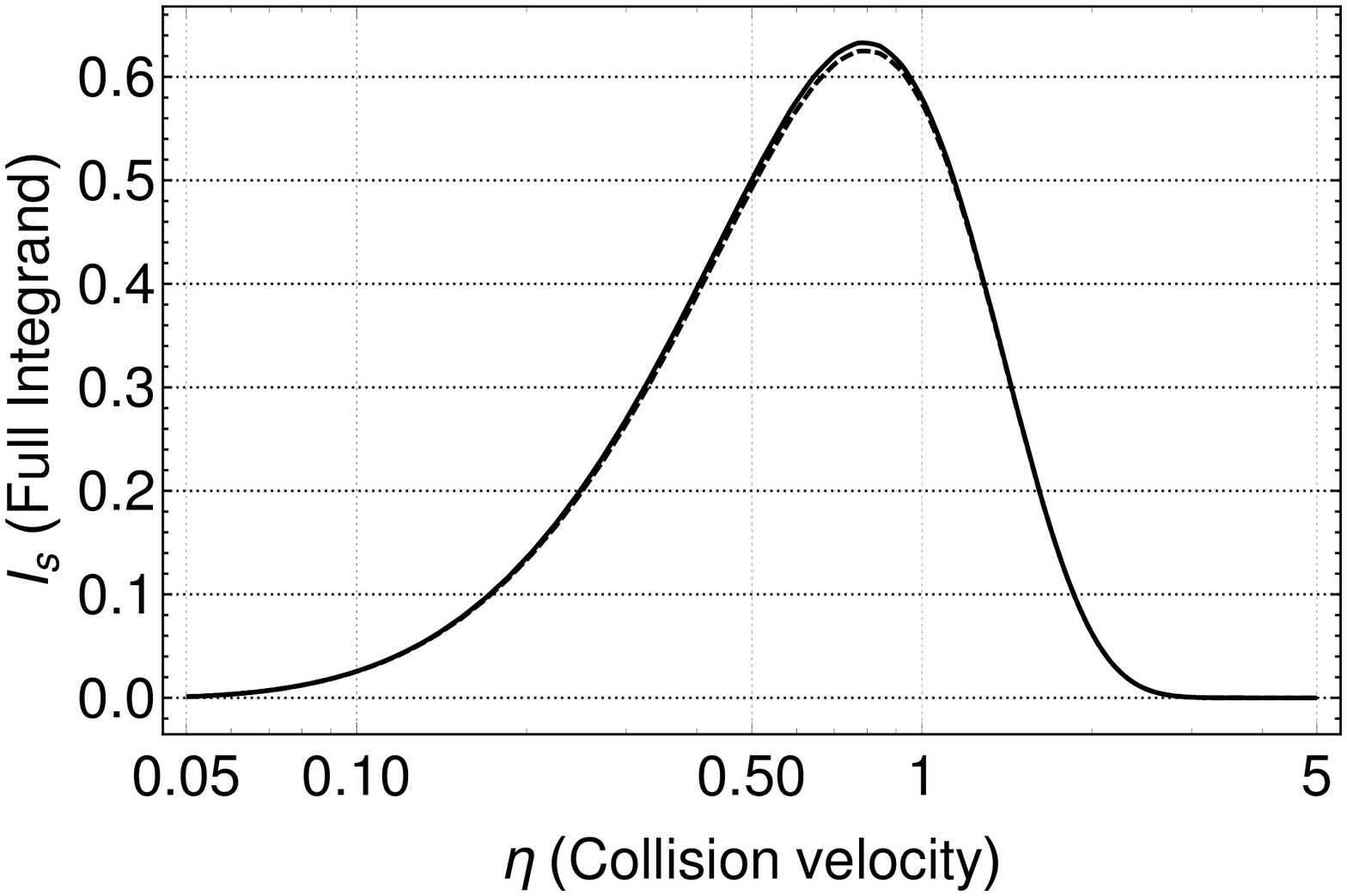}\\
\includegraphics[width=8.6cm]{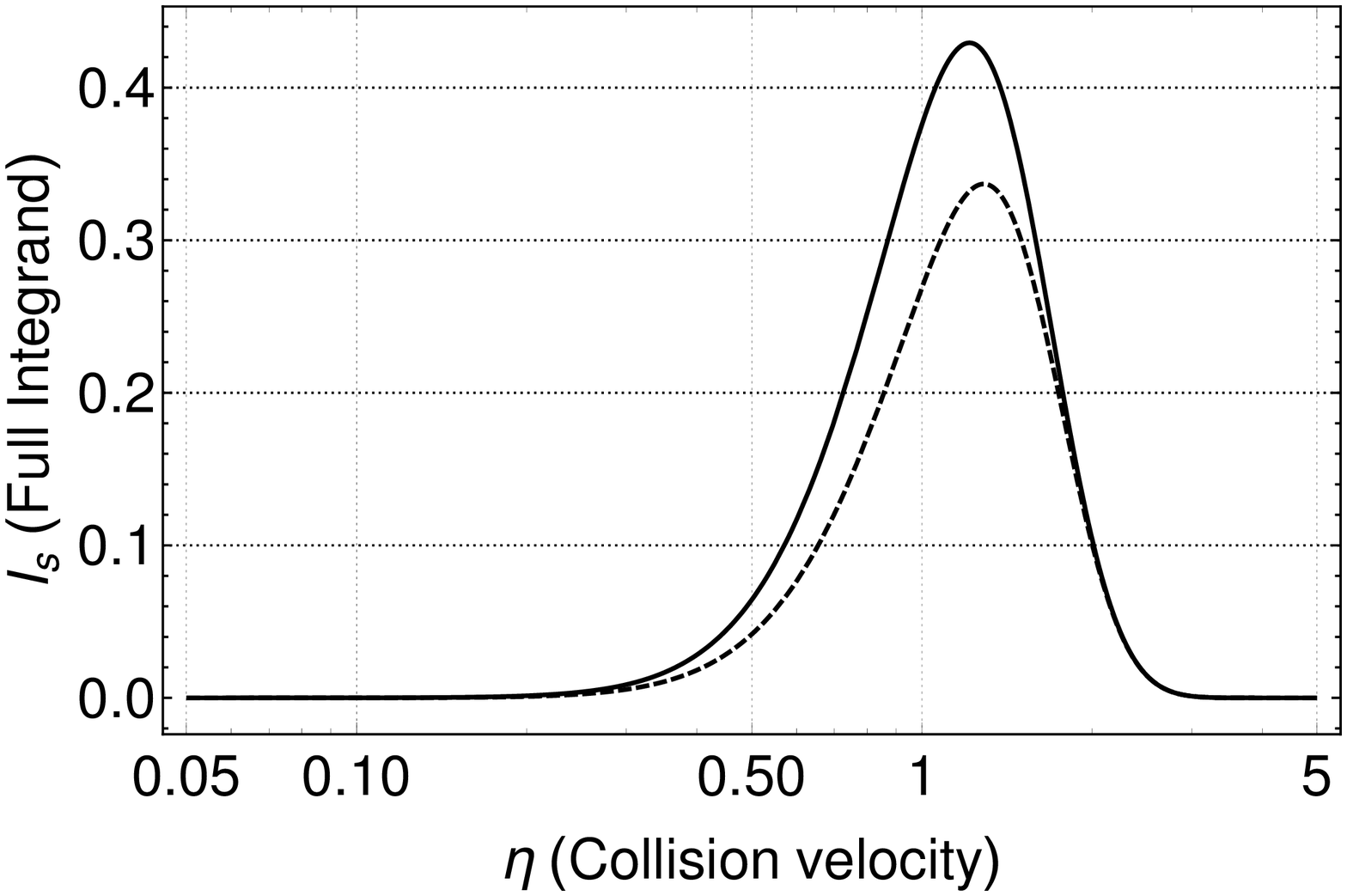}\\
\includegraphics[width=8.6cm]{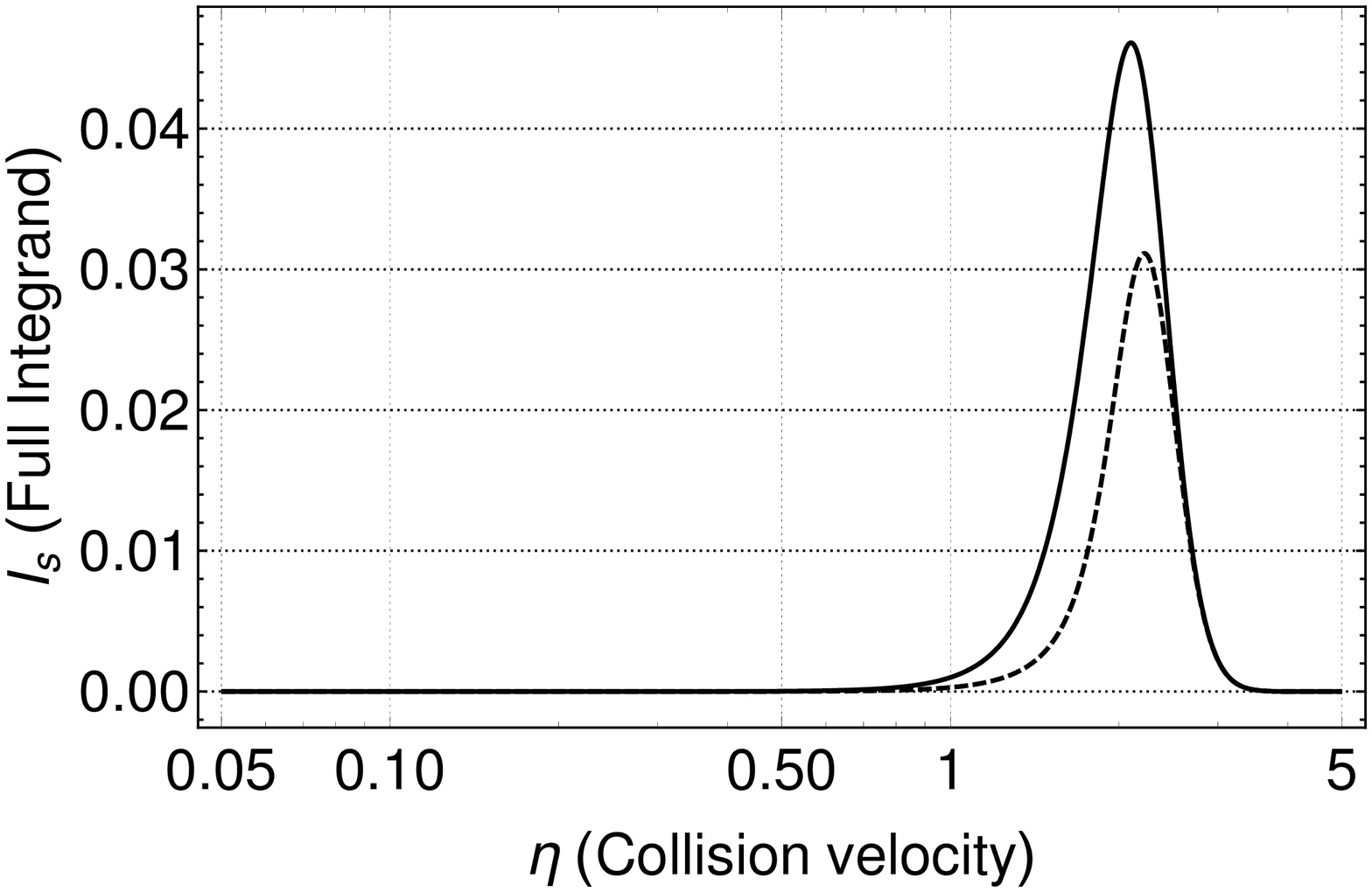}

\caption{\label{fig:momINT} Integrands ($I_{\epsilon}(\eta)$ solid and $I_{p}(\eta)$
dashed) appearing in equations (\ref{eq:xiE1}) and (\ref{eq:xiMOM1})
for deuterium at $0.774\thinspace{\rm g\cdot cm}^{-3}$ at three different
temperatures: $200\thinspace{\rm eV}$ ($\Theta\gg1$) (top), $12\thinspace{\rm eV}$
($\Theta\sim1$) (middle), $3\thinspace{\rm eV}$ ($\Theta\ll1$)
(bottom).}
\end{figure}

\section{\label{sec:discussion}Relaxation Rates in Dense Deuterium Plasma}
\begin{widetext}
We illustrate the relaxation rates predicted by the quantum mean force
theory for the case of dense deuterium with structural data provided
by the average-atom two-component-plasma model (see \citep{Starrett2012,Starrett2013}).
We compare $\sigma_{2}^{(1)}$ with $\sigma_{1}^{(1)}$ in figure
\ref{fig:momCC}. The combined influence of the negative values of
$\sigma_{2}^{(1)}$ and the preceding negative sign in equation (\ref{eq:xiMOM1})
leads to interesting behavior in the integrand for the Coulomb integral.
The full integrand of equation (\ref{eq:xiMOM1}) is shown in figure
\ref{fig:momINT} where it is seen that the resulting integrals are
positive, as required. Note that the integrands are peaked functions;
broad and peaked near the thermal velocity $v_{Ts}$ in the classical
limit, and narrow with peak near the Fermi velocity in the degenerate
limit. Also note that $I_{p}$ and $I_{\epsilon}$ are identical in
the classical limit, but differ substantially in the degenerate case
due to the presence of the $\sigma_{2}^{(1)}$ factor.
\end{widetext}

\begin{figure}[b]
\includegraphics[width=8.6cm]{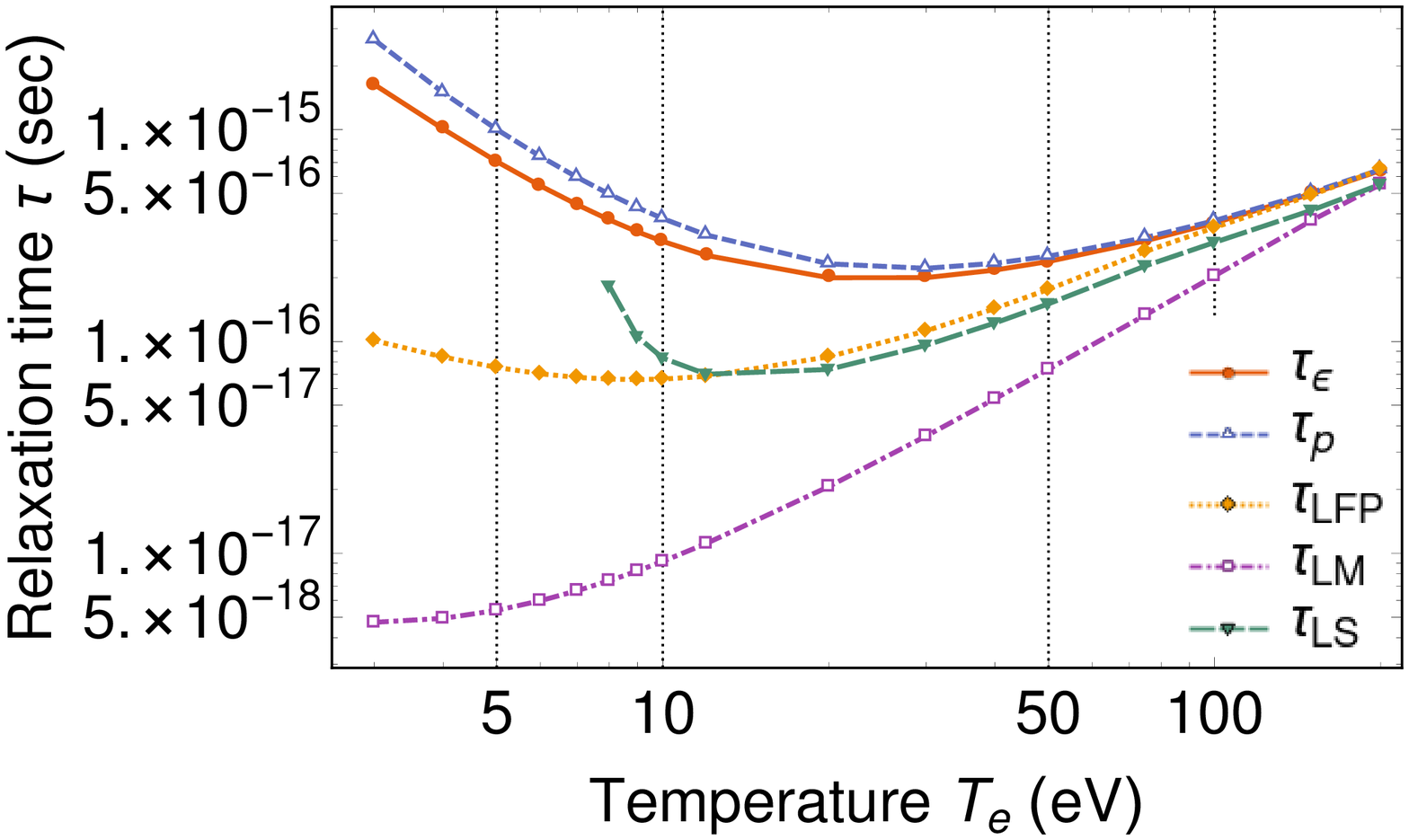}\\
\includegraphics[width=8.6cm]{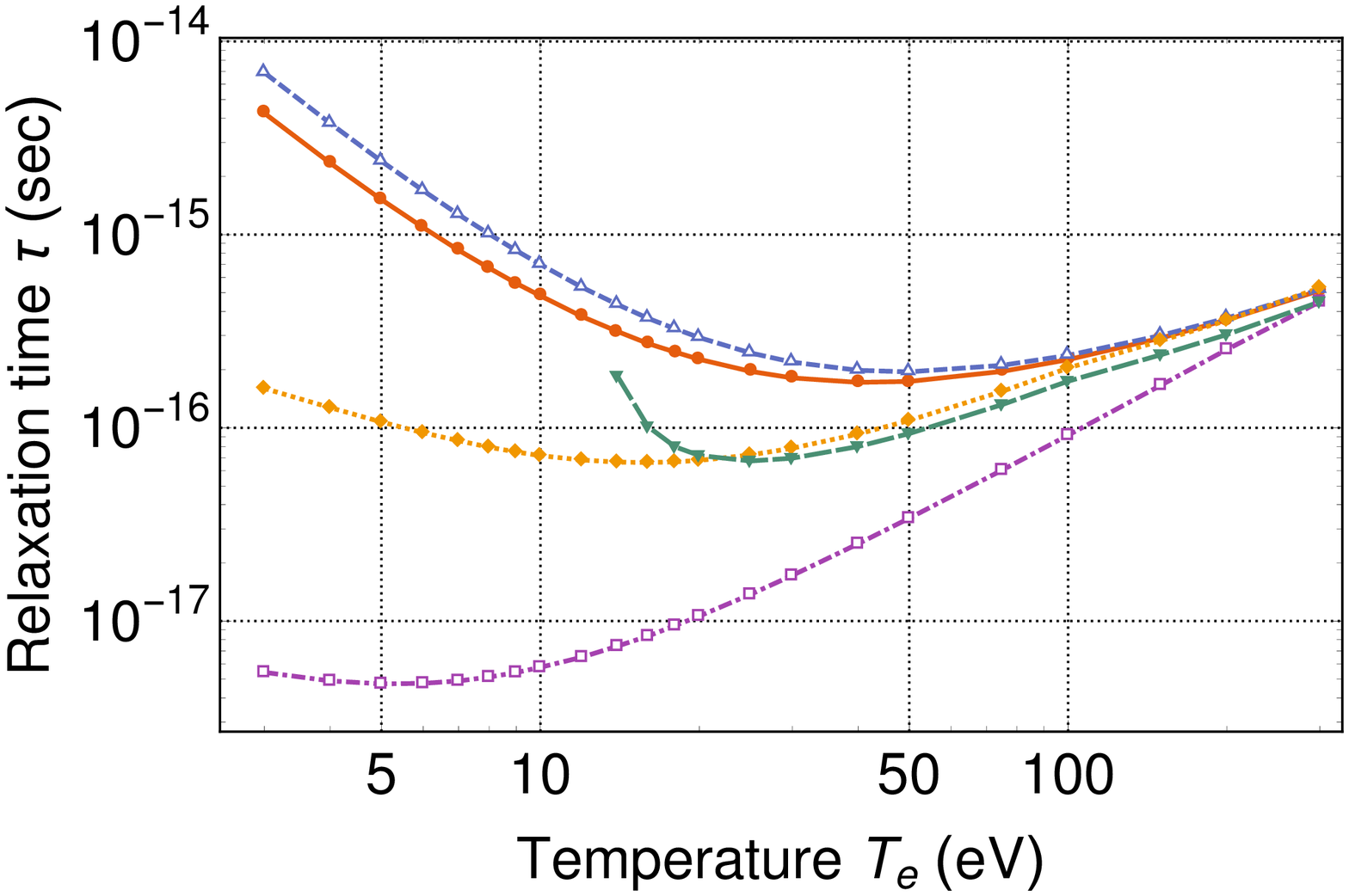}\\
\includegraphics[width=8.6cm]{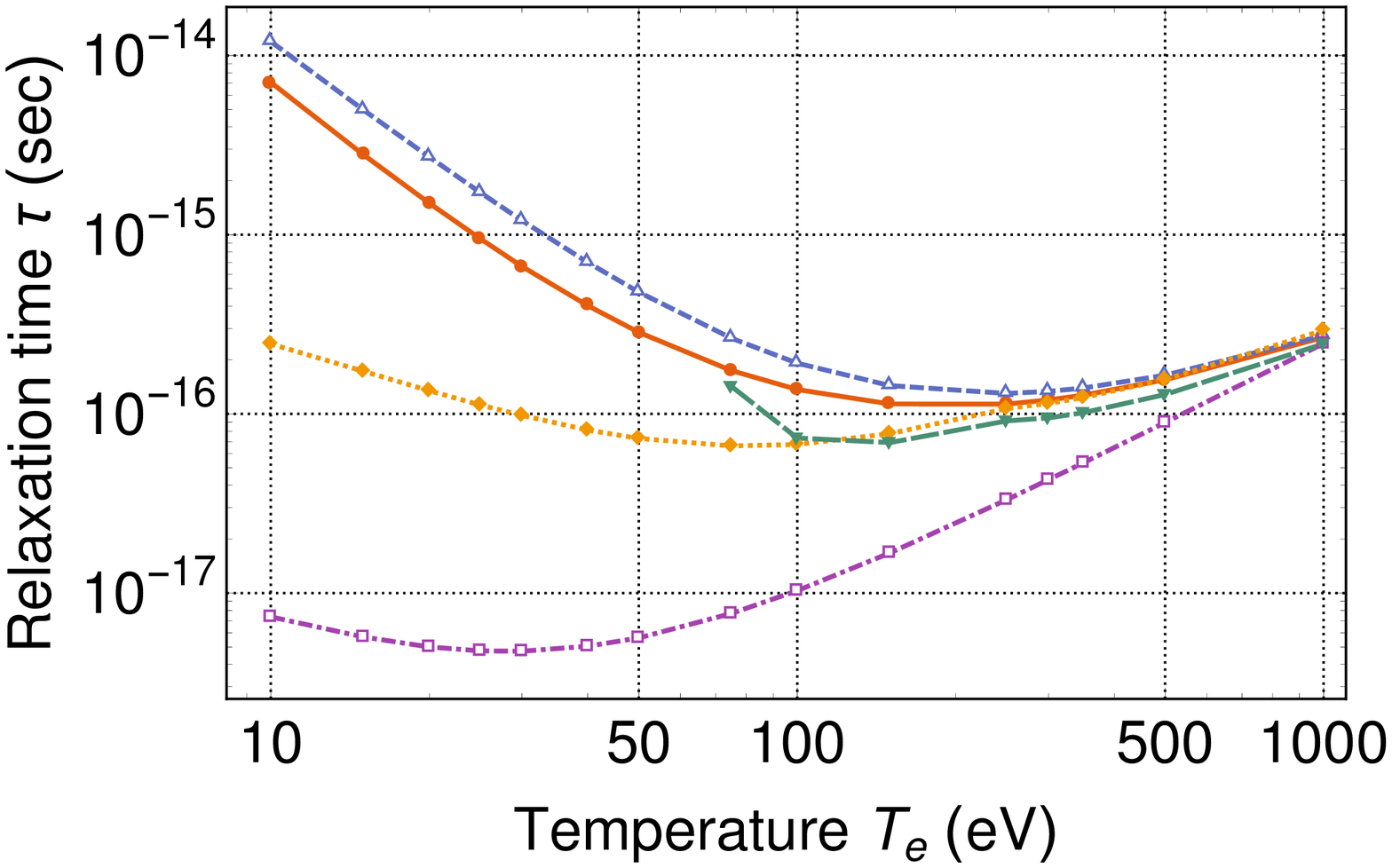}\caption{\label{fig:wide-1}Electron-ion collisional relaxation times ($\tau=\nu_{ei}^{-1}$)
as a function of temperature in deuterium at $0.774\thinspace{\rm g\cdot{\rm cm}^{-3}}$
(top), $1.75\thinspace{\rm g\cdot{\rm cm}^{-3}}$ (middle) and $20\thinspace{\rm g\cdot{\rm cm}^{-3}}$
(bottom).}
\end{figure}

\begin{figure}[b]
\includegraphics[width=8.6cm]{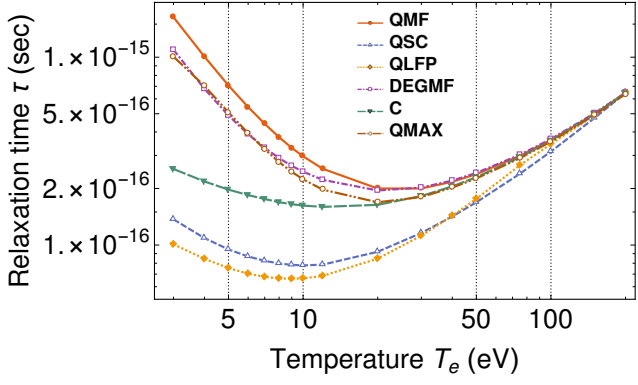}\caption{\label{fig:methodcomps}Comparison of temperature relaxation times
from differing methods in order to elucidate the relative importance
of Pauli blocking, correlations, large-angle scattering, and diffraction,
for deuterium at $0.774\thinspace{\rm g\cdot cm}^{-3}$. The lines
correspond to the different levels of evaluation of the collision
operator in an attempt to ``turn off'' different combinations of
physical processes: QMF (quantum mean-force kinetic theory), QSC (BUU
with screened potential), QLFP (quantum LFP equation {[}screened potential
with weak collisions{]}), DEGMF (QMFKT with the potential of mean
force and classical scattering), C (fully-classical mean-force theory),
and QMAX (mean-force theory with quantum scattering but Maxwellian
statistics).}
\end{figure}

\begin{figure}[b]
\includegraphics[width=8.6cm]{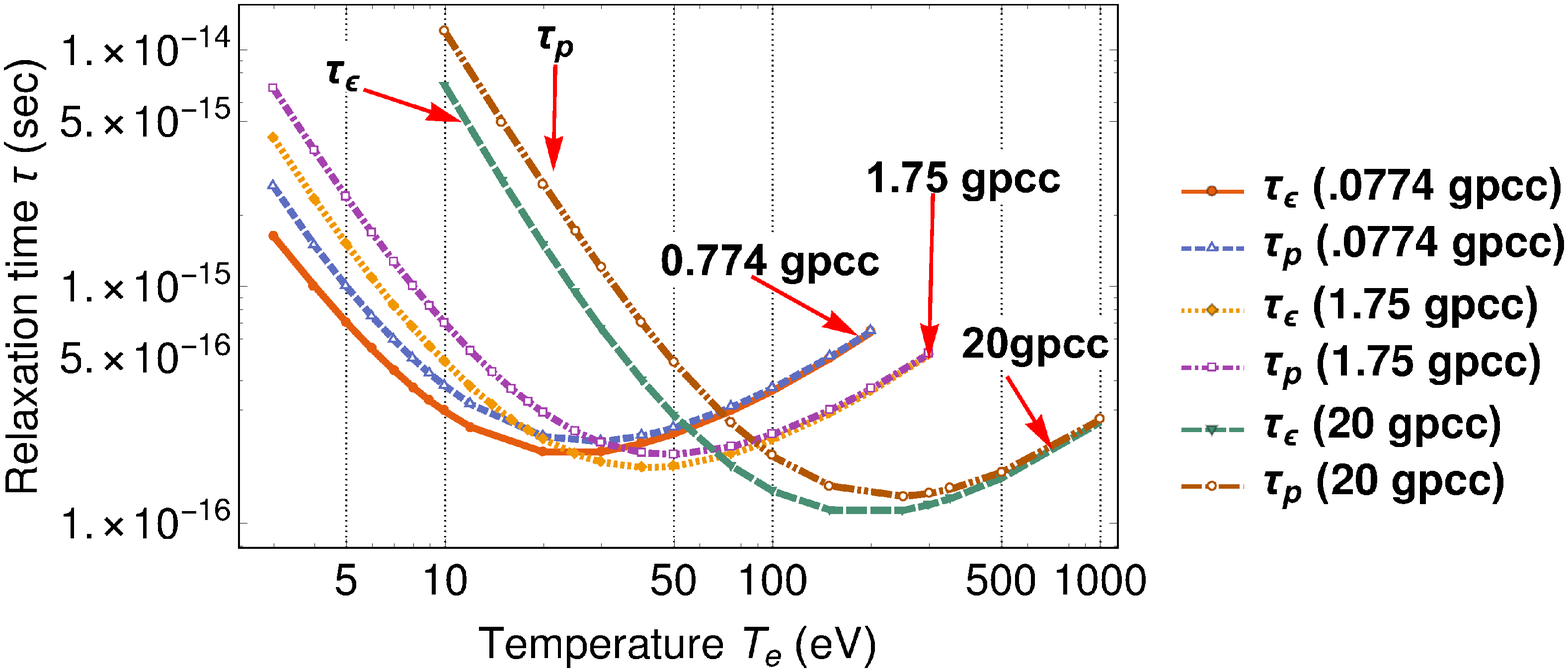}\caption{\label{fig:TPcomp}Comparison of temperature and momentum relaxation
times in all three density regimes considered. In each case rates
differ for $T<T_{F}$ but do not diverge widely even at very low temperature.
The rates are smaller for temperature relaxation than for momentum
relaxation.}
\end{figure}

We proceed to use equations (\ref{eq:Trelax}) and (\ref{eq:momrelax})
to calculate relaxation rates for the case of deuterium compressed
to $0.774$, $1.75$, and $20$ ${\rm g}\,{\rm cm}^{-3}$. So as to
explore the parameter space targeted by our model; for each density
we consider a range of temperatures so that the system ranges from
classical and weakly coupled to degenerate with moderately coupled
ions. The selected points can be seen in figure \ref{fig:parregimes}.
We compare both the temperature and momentum relaxation rates to traditional
weak-coupling models: the Landau-Spitzer rate, the quantum Landau-Fokker-Planck
(LFP) rate \citep{Daligault2016a}, and the Lee-More (LM) rate \citep{Lee1984}.
These results are shown in figures \ref{fig:wide-1}-\ref{fig:TPcomp}.

Each model reproduces the well-established Landau-Spitzer result
\begin{equation}
\nu_{ei}^{{\rm LS}}=\nu_{0}\ln\Lambda\label{eq:nuLS}
\end{equation}
in the classical weakly-coupled limit. For a given density, the models
diverge as temperature decreases and the threshold is crossed into
either strong-coupling or degenerate regimes. For reference, we provide
the scattering rates predicted by the LM and LFP theories here. The
electrical conductivity coefficient predicted by the LM model \citep{Lee1984}
is
\begin{equation}
\sigma_{e}=\frac{ne^{2}}{m}\left\{ \frac{3\sqrt{m}(kT)^{3/2}}{2\sqrt{2}\pi Z^{2}n_{i}e^{4}\ln\Lambda}\right\} \frac{4}{3}\frac{\int_{0}^{\infty}\frac{t^{3}dt}{1+\exp(t-\mu/kT)}}{\int_{0}^{\infty}\frac{t^{1/2}dt}{1+\exp(t-\mu/kT)}}
\end{equation}
which is related to the friction force density $\boldsymbol{R}$ and
thus the scattering rate: $\nu_{ei}^{(p)}=e^{2}n_{e}/\sigma m_{e}$
giving
\begin{equation}
\nu_{ei}^{{\rm LM}}=\nu_{0}\left[\ln\Lambda\frac{{\rm Li}_{3/2}(-\xi)}{{\rm Li}_{4}(-\xi)}\right].
\end{equation}
The quantum LFP equation predicts a collision rate of \citep{Daligault2016a}
\begin{equation}
\nu_{ei}^{{\rm LFP}}=\nu_{0}\left(\ln\Lambda\frac{\xi}{1+\xi}\frac{3\sqrt{\pi}\Theta^{3/2}}{4}\right).
\end{equation}
We further note our expression for the temperature relaxation rate
(given by equations (\ref{eq:Trelax})-(\ref{eq:momCC1})) is the
same as that recently obtained by a substantially different method
as equations (71)-(75) by Daligault and Simoni \citep{Daligault2019}.
This equivalency can be seen through use of the relation $n_{e}\left(h/\sqrt{\pi}m_{e}V_{e}\right)^{3}=-2{\rm Li}_{3/2}\left(-\xi\right)$
from the normalization of the Fermi-Dirac distribution. We point out
that the potential of mean force does not appear naturally in their
expression but can be included ad-hoc in the calculation of the transport
cross section.

The characteristic divergence of the Landau-Spitzer result due to
the presence of the (inverse) Coulomb logarithm
\begin{equation}
\ln\Lambda_{{\rm LS}}=\ln\frac{b_{{\rm max}}}{b_{{\rm min}}},
\end{equation}
restricts it from estimating transport even at moderate coupling.
The maximum impact parameter is modeled as the larger of either the
screening length $\lambda_{{\rm sc}}$ or the Wigner-Seitz radius
$a=(3/4\pi n_{i})^{1/3}$, and the minimum is either the classical
distance of closest approach $r_{L}=e^{2}/k_{B}T$, or more typically
in dense plasmas the thermal de Broglie wavelength $\lambda_{{\rm dB}}=\hbar/\left(m_{e}k_{{\rm B}}T_{e}\right)^{1/2}$.
In dense plasmas the vanishing Coulomb logarithm is often resolved
through the modification (see e.g. \citep{Lee1984})
\begin{equation}
\ln\Lambda_{{\rm LFP}}=\frac{1}{2}\ln\left(1+\frac{b_{{\rm max}}^{2}}{b_{{\rm min}}^{2}}\right),
\end{equation}
which we apply in the LM and LFP cases. Furthermore in the LM model
it is (artificially) enforced that the minimum value of the Coulomb
logarithm be 2 \citep{Lee1984}:
\begin{equation}
\ln\Lambda_{{\rm LM}}={\rm max}\left[2,\frac{1}{2}\ln\left(1+\frac{b_{{\rm max}}^{2}}{b_{{\rm min}}^{2}}\right)\right].\label{eq:lambdaLM}
\end{equation}
The approximations inherent in this approach are two-fold: small-angle
collisions must be assumed to obtain the LFP equation, and the choice
of maximum and minimum impact parameters represents an uncontrolled
expansion in the strongly coupled regime. The convergent kinetic equation
in our approach avoids these limitations.

Temperature relaxation times are shown in figure \ref{fig:wide-1}
for compressed deuterium at $0.774,\thinspace1.75\thinspace{\rm and}\thinspace20\thinspace{\rm g}\cdot{\rm cm}^{-3}$.
The potential of mean force force is calculated from pair distributions
obtained via the average-atom/two-component plasma method of \citep{Starrett2012,Starrett2013}.
The LS, LM and LFP rates are calculated according to equations (\ref{eq:nuLS})-(\ref{eq:lambdaLM}).
All expressions agree at high temperatures in the weakly-coupled classical
regime, while at low temperature the models differ by a factor of
$10^{2}-10^{3}$ as a result of the different levels of inclusion
of the physics of strong coupling and degeneracy. In each case there
is a minimum in the relaxation time. For the LS result this is a consequence
of the disappearance of $\ln\Lambda$, indicating the transition to
strong coupling. If the density is less than $10^{23}\thinspace{\rm cm}^{-3}$
this will occur at a temperature at which the plasma is not yet degenerate,
and if the density is greater than $10^{23}\thinspace{\rm cm}^{-3}$
the electrons will be degenerate. However, in the other theories the
minimum in $\tau$ persists, but is most prominent in the QMFKT while
being least significant in the LM and LFP theories that do not account
for strong coupling. This minimum can be attributed to a combination
of both degeneracy and strong coupling.

At high temperatures, the plasma is classical and weakly coupled,
and at low temperatures the plasma is degenerate and strongly coupled,
while in between there is a mixture of moderate coupling and degeneracy
effects. In order to assess the relative impact of large-angle collisions,
correlations, Pauli blocking, and diffraction, we compare calculations
with the following models: qLFP (no large-angle collisions or correlations),
classical MFKT (no diffraction or Pauli blocking), quantum mean-force
theory with classical scattering (no diffraction), quantum kinetic
theory with a screened Yukawa potential (neglects strong correlations)
and quantum mean-force theory with Maxwellian statistics (diffraction
but no Pauli blocking), and finally the full QMFKT with degeneracy,
diffraction, correlations and large-angle collisions included. We
illustrate this comparison in figure \ref{fig:methodcomps} for the
case of deuterium at $0.774\thinspace{\rm gpcc}$. Beginning with
the full QMFKT, we can ``turn off'' various physical effects to
determine their relative importance. For low-enough temperatures,
the Pauli blocking contribution becomes significant as there is a
large difference between the calculations with and without degeneracy,
comparable to the difference between the weakly-coupled quantum LFP
theory and the full QMFKT. The impact of diffraction continues to
higher temperatures as can be seen in comparison of the calculations
with quantum versus classical scattering. The relative importance
of correlations and strong collisions vary depending on the density
regime. At lower densities, both correlations and large-angle collisions
become important at higher temperatures than the Fermi temperature;
at higher densities strong collisions are suppressed by the Pauli
blocking at all temperatures that would otherwise be strongly coupled.
However, the correlations still persist at these densities and influence
the scattering. Notably, the correlations and Pauli blocking both
contribute to the increasing relaxation times at low temperature,
as can be noted by comparison of the classical mean-force and the
BUU with screened potential models which both flatten, with the combined
effect in the full quantum mean-force model being a large increase
in relaxation time at low temperatures. The BUU results with the screened
Coulomb potential most closely resemble the LFP results as could be
expected from the weak-coupling approximation. However, our method
allows for large-angle collisions in addition to the correlations.

The relaxation time $\tau$ for energy and momentum relaxation are
shown for compressed deuterium at three densities in figure \ref{fig:TPcomp}.
The rates are equal in the classical limit as expected, and differ
for lower temperatures when degeneracy arises. The relaxation times
differ for $T<T_{F}$, and notably the difference is larger for the
higher density cases, even for similar levels of degeneracy $\Theta$.
Generally, the rates are smaller for energy relaxation, with the maximum
difference being a factor of $\sim2$ at the lowest density $0.774\thinspace{\rm g}\cdot{\rm cm}^{-3}$
and only slightly larger at the higher densities. It is unclear whether
the collision frequencies themselves differ for the case of different
electron and ion temperatures versus different drift velocities. This
could be clarified through a more general calculation of relaxation
in a plasma with both different electron/ion temperatures and drift
velocities. Experimental validation of this phenomenon will require
accurate measurements of both momentum and temperature relaxation
rates in WDM, a matter of considerable difficulty. However, further
consideration of the physical basis for the difference between momentum
and temperature relaxation is called for, and perhaps computational
methods will prove to be effective to this end.

\section{\label{sec:conclusions}Conclusions}


We have presented a model for transport in plasmas with weak to moderate
Coulomb coupling and weak to moderate electron degeneracy. The model
is derived from the quantum BBGKY hierarchy with a closure scheme
that expands about the deviation of correlations from their equilibrium
values rather than in terms of the strength of the correlations. This
incorporates correlations in the equilibrium limit while maintaining
the simplicity of binary collisions in the dynamical equation. Although
the expansion provides a general framework, it is difficult to solve
in practice. A practical model was obtained by applying a quasi-classical
approximation, closure at $n=2$, and the molecular chaos approximation.
This is relevant to electron-ion collisions in warm dense matter.
The result has the form of a Boltzmann-Uehling-Uhlenbeck equation,
but where the binary collisions are mediated by the electron-ion potential
of mean force (PMF), which is calculated directly from the pair distribution
function $g_{ei}(r).$ Any means of obtaining $g_{ei}$ thus provides
a means of obtaining the PMF. 

The model was used to evaluate relaxation rates in warm dense deuterium,
in which the $g_{ei}$ were provided by an average-atom two-component-plasma
hybrid model \citep{Starrett2012}. The temperature relaxation rate
was written analogously to the classical Landau-Spitzer (LS) result
in terms of a ``Coulomb integral'' that takes the place of the traditional
Coulomb logarithm. The Coulomb integral for temperature relaxation
depends on the level of degeneracy, and Coulomb coupling enters through
the calculation of the momentum-transfer cross section solving the
Schr\"{o}dinger equation with the PMF as the scattering potential.
The momentum relaxation rate differs from temperature relaxation in
that it depends on a different transport cross section, which includes
a term that is solely associated with degeneracy, and has no analog
in the classical limit. The dependence of the integrands of the Coulomb
integrals on the level of degeneracy was compared for the temperature
and momentum relaxation cases.

We concluded by calculating the temperature and momentum relaxation
rates in dense deuterium at three different densities over temperature
ranges to cover the transitions between weak and moderate coupling
and weak and moderate degeneracy. The temperature and momentum relaxation
rates were compared with each other, and also compared with other
common approximations for the relaxation rates. It was found that
all models agree in the classical weak-coupling limit as expected,
and diverge widely in the limit of a degenerate moderately-coupled
plasma. We assessed the relative importance of the different relevant
physical processes that complicate the problem as degeneracy and coupling
simultaneously increase: diffraction, Pauli blocking, correlations,
and large-angle scattering. Interestingly, in the degenerate regime
there is a quantitative difference in the predicted relaxation rates
for momentum versus energy. Ultimately, current and near-future experimental
measurements and ab-initio simulations will be able to shed light
on the applicability of the different models of transport for dense
plasmas and WDM.

Further extensions of this work may be possible by solving the general
formulation, rather than the semiclassical limit, to address electron-electron
collisions in WDM. In this case the definition and interpretation
of the PMF is complicated by both the exclusion and uncertainty principles.
Further work will be required to obtain a rigorously derived convergent
kinetic equation for electron-electron collisions. Finally, recent
and upcoming experimental measurements of electrical conductivity
and temperature relaxation \citep{Cho2016,Zaghoo2019} may soon open
the door for discrimination between the validity of the various models
of relaxation in dense plasmas. This will enhance our understanding
of the basic physics of dense plasmas, and allow increased fidelity
in the rapid calculation of transport coefficients for use in hydrodynamic
simulations of naturally and experimentally occurring WDM.
\begin{acknowledgments}
The authors wish to acknowledge Charles Starrett for the provision
of input data at equilibrium for the potential of mean force. This
material is based upon work supported by the U.S. Department of Energy,
Office of Science, Office of Fusion Energy Sciences under Award Number
DE-SC0016159.
\end{acknowledgments}

\appendix

\section{\label{sec:wignerkirkwoodappendix}Wigner Functions in The Wigner-Kirkwood
Expansion}

The reduced Wigner functions can be expressed in the semi-classical
Wigner-Kirkwood expansion \citep{Shalitin1973} as
\begin{gather*}
f_{0}^{(n)}=\exp\left(-\frac{\beta}{2m}\sum_{i}^{n}p_{i}^{2}\right)\int d\boldsymbol{r}^{(N-n)}\exp\left(-\beta U\right)\\
\times\Bigg\{1+\hbar^{2}\beta^{2}\Bigg[-\frac{1}{24m}\sum_{i>n}\frac{\partial^{2}U}{\partial\boldsymbol{r}_{i}^{2}}\\
-\frac{1}{8m}\sum_{i\leq n}\frac{\partial^{2}U}{\partial\boldsymbol{r}_{i}^{2}}+\frac{\beta}{24m}\sum_{i\leq n}\left(\frac{\partial U}{\partial\boldsymbol{r}_{i}}\right)^{2}\\
+\frac{\beta}{24m^{2}}\sum_{i_{1}<n}\sum_{i_{2}\leq n}\frac{\partial^{2}U}{\partial\boldsymbol{r}_{i_{1}}\partial\boldsymbol{r}_{i_{2}}}:\boldsymbol{p}_{i_{1}}\boldsymbol{p}_{i_{2}}\Bigg]\Bigg\},
\end{gather*}
in which \textbf{$\beta\equiv1/k_{B}T$}. Defining
\begin{gather*}
C_{n}=\exp\left(-\frac{\beta}{2m}\sum_{i}^{n}p_{i}^{2}\right)\int d\boldsymbol{r}^{(N-n)}\exp\left(-\beta U\right)\\
Q_{n}=\hbar^{2}\beta^{2}\exp\left(-\frac{\beta}{2m}\sum_{i}^{n}p_{i}^{2}\right)\int d\boldsymbol{r}^{(N-n)}\exp\left(-\beta U\right)\\
\times\Bigg[-\frac{1}{24m}\sum_{i>n}\frac{\partial^{2}U}{\partial\boldsymbol{r}_{i}^{2}}-\frac{1}{8m}\sum_{i\leq n}\frac{\partial^{2}U}{\partial\boldsymbol{r}_{i}^{2}}+\frac{\beta}{24m^{2}}\sum_{i\leq n}\left(\frac{\partial U}{\partial\boldsymbol{r}_{i}}\right)^{2}\\
+\frac{\beta}{24m^{2}}\sum_{i_{1}<n}\sum_{i_{2}\leq n}\frac{\partial^{2}U}{\partial\boldsymbol{r}_{i_{1}}\partial\boldsymbol{r}_{i_{2}}}:\boldsymbol{p}_{i_{1}}\boldsymbol{p}_{i_{2}}\Bigg]
\end{gather*}
we can write
\begin{equation}
\frac{f_{0}^{(3)}}{f_{0}^{(2)}}=\frac{C_{3}+Q_{3}}{C_{2}+Q_{2}}
\end{equation}
and, taking the expansion terms $Q_{n}$ to be small compared to the
$C_{n}$, obtain 
\begin{equation}
\frac{f_{0}^{(3)}}{f_{0}^{(2)}}\approx\exp\left(-\frac{\beta}{2m}p_{3}^{2}\right)\frac{g_{3}}{g_{2}}\left[1+\left(\frac{Q_{3}}{C_{3}}-\frac{Q_{2}}{C_{2}}\right)\right],
\end{equation}
where $g_{3}=\int\exp\left(-\beta U\right)d^{3(N-3)}r$. The term
in parenthesis is the first-order quantum correction, it is suppressed
for small coupling and for small degeneracy. For the case of electron-ion
collisions it can be neglected if both the degeneracy and coupling
are only small to moderate. Dropping this term we obtain equation
(\ref{eq:wignerapprox}).

\section{\label{sec:shifts_appendix}Simplification of Cross Sections}

Cross sections are calculated in the partial wave expansion
\begin{equation}
\frac{d\sigma}{d\Omega}=\left|\frac{1}{2ik}\sum_{l=0}^{\infty}\left(2l+1\right)\left({\rm e}^{2i\delta_{l}}-1\right)P_{l}\left({\rm cos}\theta\right)\right|^{2},
\end{equation}
where the phase shifts $\delta_{l}$ are calculated from solution
of the Schr\"{o}dinger equation for the given potential. This can
be written as a double sum
\begin{gather*}
\frac{d\sigma}{d\Omega}=\frac{1}{k^{2}}\sum_{n=0}^{\infty}\sum_{m=0}^{\infty}\left(2m+1\right)\left(2n+1\right)\\
\times{\rm e}^{i\left(\delta_{m}-\delta_{n}\right)}\sin\delta_{m}\sin\delta_{n}{\rm P}_{m}(\cos\theta){\rm P}_{n}(\cos\theta)
\end{gather*}
which inserted into equation (\ref{eq:momCC2a}) results in
\begin{gather*}
\sigma_{2}^{(1)}\left(\eta,\Gamma\right)=\frac{4\pi}{k^{2}}\\
\times\sum_{n=0}^{\infty}\sum_{m=0}^{\infty}\left(2m+1\right)\left(2n+1\right){\rm e}^{i\left(\delta_{m}-\delta_{n}\right)}\sin\delta_{m}\sin\delta_{n}\\
\times\int_{0}^{\pi}d\theta\sin^{2}\frac{\theta}{2}\sin\theta\cos\theta{\rm P}_{m}(\cos\theta){\rm P}_{n}(\cos\theta).
\end{gather*}
Taking advantage of the properties of the Legendre polynomials this
is
\begin{gather*}
\frac{4\pi}{k^{2}}\sum_{n=0}^{\infty}\sum_{m=0}^{\infty}\left(2m+1\right)\left(2n+1\right){\rm e}^{i\left(\delta_{m}-\delta_{n}\right)}\sin\delta_{m}\sin\delta_{n}\\
\times\frac{1}{2}\int_{0}^{\pi}\sin\theta\left[P_{1}(\cos\theta)-\frac{2}{3}P_{2}(\cos\theta)-\frac{P_{0}}{3}\right]\\
\times P_{m}(\cos\theta)P_{n}(\cos\theta)d\theta,
\end{gather*}
Finally, with the identity
\begin{equation}
\int_{0}^{\pi}d\theta\sin\theta P_{l}(\cos\theta)P_{m}(\cos\theta)P_{n}(\cos\theta)=2\left(\begin{array}{ccc}
l & m & n\\
0 & 0 & 0
\end{array}\right)^{2},
\end{equation}
where $\left(\begin{array}{c}
\cdots\\
\cdots
\end{array}\right)$ is the Wigner 3j symbol and using the specific values for $l=0,\thinspace1,\thinspace{\rm and}\thinspace2$,
we obtain equation (\ref{eq:momCC2b}).

\bibliographystyle{apsrev4-1}
\bibliography{RightleyBaalrud_PRE_2020_01_29}

\end{document}